\documentstyle[12pt]{article}
\begin{document}

\tolerance=5000

\def\pp{{\, \mid \hskip -1.5mm =}}
\def\cL{{\cal L}}
\def\be{\begin{equation}}
\def\ee{\end{equation}}
\def\bea{\begin{eqnarray}}
\def\eea{\end{eqnarray}}
\def\tr{{\rm tr}\, }
\def\nn{\nonumber \\}
\def\e{{\rm e}}
\def\D{{D \hskip -3mm /\,}}

\  \hfill
\begin{minipage}{3.5cm}
UPR-1044-T  \\
\end{minipage}

\vfill

\begin{center}
{\large\bf Cosmological  anti-deSitter space-times  and
time-dependent AdS/CFT correspondence}

\vfill

{\sc Mirjam CVETI\v C}\footnote{cvetic@cvetic.hep.upenn.edu. On
sabbatic leave from the  University of Pennsylvania.},\\
{\sc Shin'ichi NOJIRI}$^\clubsuit$\footnote{snojiri@yukawa.kyoto-u.ac.jp, 
nojiri@cc.nda.ac.jp}
and {\sc Sergei D. ODINTSOV}$^{\spadesuit}
$\footnote{odintsov@ieec.fcr.es, 
odintsov@mail.tomsknet.ru.
Also at Tomsk State Pedagogical University, Tomsk, RUSSIA.}
\vfill

{\sl School of Natural
Sciences, Institute for Advanced Studies,\\
 Princeton NJ 08540, USA}

\vfill

{\sl $\clubsuit$ Department of Applied Physics \\
National Defense Academy,
Hashirimizu Yokosuka 239-8686, JAPAN}

\vfill

{\sl $\spadesuit$
ICREA and IEEC, \\
Edifici Nexus, Gran Capit\`a 2-4, 08034, Barcelona, SPAIN}


\begin{abstract}
We study classes of five-dimensional cosmological solutions with
negative curvature, which  are  obtained from static solutions
by an exchange of a spatial and  temporal coordinate, and in some cases
by  an analytic
continuation. Such solutions provide a suitable
laboratory to address the time-dependent AdS/CFT correspondence.
For a specific example we address in detail the calculation of the
boundary stress-energy and the Wilson line and
find disagreement with the standard AdS/CFT correspondence.
We trace these discrepancies  to  the time-dependent effects,
such as particle creation, which we further study
for specific backgrounds.
We also  identify    specific time-dependent
backgrounds  that
reproduce the correct conformal anomaly. For such backgrounds the
calculation of the Wilson
line in the adiabatic approximation  indicates only
a Coulomb repulsion.
\end{abstract}
 \end{center}

\vfill


\newpage

\section{Introduction}

The interest in  time-dependent backgrounds in  string theory is
motivated  by  several reasons. The most important one is related to
the attempts to
understand the  origin of the very early Universe, in particular within
superstring   theory  and its M-theory unification, which are the most
promising  candidate for explaining the  underlying unification of
forces of nature. If string theory  were able to shed light on  the
aspect of the early Universe, this would provide a strong incentive for
its further investigation.
 From another point of view,   the study of such time-dependent
backgrounds  may  provide new insights into
underlying symmetries that govern   cosmological evolutions.
For instance, AdS/CFT duality relates the properties of string
theory on negative curvature D-dimensional space-times to the dual
(D-1)-dimensional supersymmetric gauge theory.
A time-dependent version of such a duality may in turn relate the
properties of   time-dependent gravitational evolutions
(potentially describing the early
Universe) to a dual description in terms  of gauge
theories in one dimension less.

It is known that quantum field theory (QFT)  in curved time-dependent
space-time
(see e.g., \cite{BOS} for introduction) is much more involved than the
corresponding theory in a static space. In particular, there occurs
particle creation caused by the gravitational field and the vacuum state
changes with time. There are
 various  non-zero quantum averages for ''in-in'', ''out-in'',
''in-out'' and ''out-out'' amplitudes.
The standard QFT is
well-developed to study only
out-in amplitudes. At the same time, in quantum cosmology one needs
in-in
quantities as well. Moreover, we expect that a formulation of the
time-dependent AdS/CFT correspondence should  encode  information
for various vacuum states in
time-dependent backgrounds.

The purpose of this paper is to discuss  aspects of (conjectured)
AdS$_5$/CFT$_4$ correspondence and its various predictions and
manifestations  when applied to  the situation when the
five-dimensional space-time  is a cosmological AdS. We obtain such
cosmological 5d AdS spaces from their AdS black hole cousins by
exchanging the spatial and temporal coordinates, along with the
corresponding analytical continuation when it is required.
Examples of such cosmological AdS spaces include  curved  branes
with deSitter space world-volume or a product of a  sphere  and a
torus. For this class of backgrounds we focus on the  evaluation
of the
 surface stress tensor and the
Wilson loop calculation on the  gravitational side. In particular,
for  these backgrounds one reproduces the holographic trace
anomaly  that agrees  with the prediction of the AdS/CFT
correspondence. On the other hand, we also demonstrate   a
discrepancy  in the calculation of the  energy  momentum tensor
components on the gravity side with that of the  AdS/CFT
predictions can be traced to the omitted non-local effects of
particle creation in the bulk.

The paper is organized as follows. In the next section it is shown how
one can construct a cosmological AdS space from the 5d Schwarzschild-AdS
black hole by
exchanging the role of temporal and radial coordinates.
This method  can also be applied to BPS solutions which appear in 5d
gauged supergravity, and we demonstrate that on a specific example of
bent BPS domain walls with an asymptotic AdS  space-time.

Section \ref{s3} is devoted to a general consideration, addressing which
5d cosmological AdS space may lead to the correct holographic conformal
anomaly.  An explicit example of such a space is presented.  As an
additional check the thermodynamical energy is found.  We  confirm that
cosmological AdS  backgrounds,  described in the previous section,
belong to the class of backgrounds with the correct  holographic anomaly.

In section \ref{s4}  we study the massive scalar propagator in a  cosmological
AdS space and demonstrate that  in these time-dependent backgrounds
 one  reproduces the conformal
dimensions of these fields, in agreement with the
 usual AdS/CFT predictions.

 In section \ref{s5}  we address  particle creation in these time-dependent
 backgrounds,  which should  provide non-local effects that account for the discrepancy
 between the standard evaluation of the boundary stress energy  on the gravity side and the
 dual field theory prediction.
In particular, we address  the  number of particles created on deSitter brane 
in a time-dependent AdS background. 

Section \ref{s6} is devoted to the
evaluation of  an  analog of
Wilson loop in the  adiabatic approximation for the cosmological AdS
spaces with  the  correct holographic conformal anomaly.
It is demonstrated that  the potential has a typically Coulomb-like form
(as for the pure AdS space).

In the last section we summarize the results and  discuss further directions.

In the Appendix  we  also summarize the
construction of  cosmological space-times that are obtained
 from the AdS  spaces by applying
$T$-duality  (within the  Neveu-Schwarz-Neveu-Schwarz sector of string theory).
In this case the  obtained space-times are
not  asymptotically AdS, but
  correspond to the so-called dilatonic  vacua, and thus the standard
 AdS/CFT correspondence is not applicable.    Nevertheless there may be remnants
 of the  standard AdS/CFT correspondence  surviving due to the fact that
  these backgrounds are
 T-dual to the  AdS  ones. With this in mind, analogs of the
 surface stress-tensor and
Wilson loop  on the gravitational side are calculated for the
T-dual backgrounds.

\section{Cosmological AdS spaces \label{s1}}

In this section we demonstrate on a number of examples how to obtain
cosmological AdS space-times from the  static ones via exchanges of  spatial
and temporal coordinates, and in some cases via further analytic continuation.
We start with an  example of  the Schwarzschild-AdS (SAdS) black hole metric:
\be
\label{AdSS}
ds_{\rm AdS-S}^2 = -f(r) dt^2 +  {dr^2 \over f(r)}
+ r^2 d\Omega_3^2\ ,\quad f(r)\equiv {r^2 \over l^2} + 1 - {\mu \over r^2}
\ee
This space-time may correspond to a classical solution of a gauged supergravity
theory that arises as a sphere  compactification of an effective supergravity
from string or M-theory.
In (\ref{AdSS}), $d\Omega_3^2$ is the metric of the three dimensional sphere
S$^3$ with unit radius. The horizon radius $r_H$ is
\be
\label{abrh1}
r_H^{2}= - {l^2 \over 2} + {1\over 2}\sqrt{ l^4 + 16 \mu l^2 }
\ee
and the Hawking temperature $T_H$
\be
\label{hwk}
T_H = \left| \left.{1\over 4\pi}{d f(r)\over da}\right|_{r=r_H}\right|
= {1 \over 2\pi r_H} + {r_H \over \pi l^2} \ .
\ee

In the following we shall discuss how to obtain the cosmological AdS
space from the above SAdS black hole.
One approach was  employed  in  ref.\cite{Aharony} (see also
\cite{BR})  writing the metric $d\Omega_3^2$ in
(\ref{AdSS}) as
\be \label{AFN12}
d\Omega_3^2 = d\chi^2 + \sin^2 \chi \left(d\theta^2 + \sin^2\theta d\phi^2\right)\ ,
\ee
and replacing
\be
\label{AFN13}
t\to i\tilde\phi \ ,\quad \chi\to {\pi \over 2} + i\tilde t\ ,
\ee
the bubble solution has been constructed:
\footnote{In the typical case of the AdS/CFT correspondence, the
space-time is given by 
AdS$_5 \times $S$_5$. Such a solution exists since there are
(self-dual five-form) fluxes in S$_5$.
Since the 
analytic continuation (\ref{AFN13}) corresponds 
to the deformation of the AdS$_5$, 
in most of cases 
the five-form flux remains  real after analytic continuation. 
For more details, see the discussion at the end of this section. 
}
\be
\label{AFN14}
ds^2 = -r^2 d\tilde t^2 + f(r)d\tilde \phi^2 + {1 \over f(r)}dr^2
+ r^2 \cosh^2 \tilde t \left(d\theta^2 + \sin^2\theta d\phi^2\right)\ .
\ee
If $\tilde\phi$ is periodic, $r$ is restricted by
\be
\label{AFN10}
r\geq r_H
\ee
and the singularity at $r=r_H$ does not appear in (\ref{AFN14}).

There is, however,  more freedom in the choice of an analytic
continuation.
For example, instead of (\ref{AFN14}), if we consider,
\be
\label{OT1}
t\to i\tilde\phi \ ,\quad \theta\to {\pi \over 2} + i\tilde t\ ,
\ee
one obtains
\be
\label{OT2}
ds^2 = -r^2 \sin^2 \chi d\tilde t^2 + f(r)d\tilde \phi^2 + {1 \over f(r)}dr^2
+ r^2 \sin^2 \chi \cosh^2 \tilde t d\phi^2\ .
\ee
Alternatively, with
\be
\label{OT3}
t\to i\tilde\phi \ ,\quad \chi \to i\tilde t\ ,\quad \theta\to i\tilde \theta\ ,
\ee
we obtain
\be
\label{OT4}
ds^2 = -r^2 d\tilde t^2 + f(r)d\tilde \phi^2 + {1 \over f(r)}dr^2
+ r^2 \sinh^2 \tilde t \left(d\tilde\theta^2 + \sinh^2 \tilde \theta d\phi^2\right)\ .
\ee
Furthermore, taking
\be
\label{OT5}
t\to i\tilde\phi \ ,\quad \phi\to i\tilde t\ ,
\ee
we obtain the static space-time:
\be
\label{OT6}
ds^2 = -r^2 \sin^2 \chi \sin^2 \theta d\tilde t^2 + f(r)d\tilde \phi^2 + {1 \over f(r)}dr^2
+ r^2 \left(d\chi^2 + \sin^2 \chi d\theta^2 \right)\ .
\ee
For (\ref{OT2}), (\ref{OT4}), or (\ref{OT6}), if $\tilde\phi$ is periodic
in (\ref{AFN9}), $r$ is restricted by (\ref{AFN10}) and
there is no singularity at $r=r_H$.

Following  \cite{NOcosm}, another cosmological model can be obtained from
the SAdS black hole.
Inside the horizon $r<r_H$ in (\ref{AdSS}), if we rename
$r$ as $t$ and $t$
as $r$ we get a metric describing a time-dependent background:
\be
\label{vii}
ds^2= - {l^2 t^2 \over \left( r_H^2 - t^2 \right)
\left(r_B^2 + t^2 \right)} dt^2
+{\left( r_H^2 - t^2 \right)
\left(r_B^2 + t^2 \right) \over l^2 t^2}dr^2
+ t^2 d\Omega_3^2 \ .
\ee
Here
\be
\label{AFN11}
r_B^2={l^2 \over 2}\left(1 + \sqrt{
1 + {4\mu \over l^2}}\right) \ .
\ee
The appearance of the time-dependent background behind the horizon of SAdS
black hole, where   the physical role of time and radial
coordinates was exchanged, is of course very analogous to the space-time
properties of regular Schwarzschild black holes behind the horizon.
Since $t=0$ corresponds to the singularity of
the black hole, there is a curvature singularity, where the
square of the Riemann tensor diverges as
\be
\label{Ri}
R_{\mu\nu\rho\sigma}R^{\mu\nu\rho\sigma}\sim {72 \over t^8}\ .
\ee
The singularity at $t=0$ might be regarded as a big-bang  singularity.
The topology of the space is R$_1\times$S$_3$, where
R$_1$ corresponds to $r$. The metric of R$_1$ vanishes at the
horizon $t=r_H$. The scalar
curvature $R$ is a  negative constant
\be
\label{x}
R=-{20 \over l^2}\ .
\ee
as it should be for a  typical  AdS space-time.

If we change the coordinate by
\be
\label{t1}
t^2 = {r_H^2 + r_B^2 \over 2}\cos {2\tau \over l}
+ {r_H^2 - r_B^2 \over 2}
= l^2\sqrt{1 + {4\mu \over l^2}}\cos {2\tau \over l}- l^2\ ,
\ee
one can rewrite the metric (\ref{vii}) in the following form:
\bea
\label{t2}
ds^2 &=& - d\tau^2 + {\left(1 + {4\mu \over l^2}\right)
\sin^2 {2\tau \over l} \over \sqrt{1 + {4\mu \over l^2}}\cos {2\tau \over l}
- 1}dr^2 \nn
&& + \left(  l^2\sqrt{1 + {4\mu \over l^2}} \cos {2\tau \over l}- l^2\right)d\Omega_3^2 \ .
\eea
In (\ref{t2}), $\sin{2\tau \over l}=0$ corresponds to the
horizon and $\cos {2\tau \over l} = {k \over 2\sqrt{1 + {4\mu \over l^2}}}$
corresponds to the curvature singularity.

Let us also point out another analytic continuation of the metric
(\ref{t2}).
First we write $d\Omega_3^2$ as in (\ref{AFN12})
and consider the following analytic continuation
\be
\label{FF2}
\tau \to i\tilde r \ ,\quad r \to i\tilde\phi \ ,\quad \phi \to i \tilde\tau\ .
\ee
Then the metric  (\ref{t2}) can be rewritten as a static one:
\bea
\label{FF3}
ds^2 &=& d\tilde r^2 + {\left(1 + {4\mu \over l^2}\right)
\sinh^2 {2\tilde r \over l} \over \sqrt{1 + {4\mu \over l^2}}\cosh {2\tilde r \over l}
- 1}d\tilde \phi^2 \nn
&& + \left(  l^2\sqrt{1 + {4\mu \over l^2}} \cosh {2\tilde r \over l}- l^2\right)
\left( d\chi^2 + \sin^2 \chi \left(d\theta^2 - \sin^2 \theta d\tilde \tau^2\right)\right) \ .
\eea
As we will see in the next section, AdS/CFT is applied to such metric.
This metric (\ref{FF3}) is, however, another presentation of the metric (\ref{OT6}),
which are related by the coordinate transformation given by
\be
\label{FF4}
r^2 = l^2\sqrt{1 + {4\mu \over l^2}}\cosh {2\tilde r \over l}- l^2\ ,
\ee
When $- C \equiv {4\mu \over l^2} <0$, a non-trivial analytic continuation
is given by
\be
\label{FFF1}
r=\tilde \tau\ ,\quad \tau = {\pi l \over 4} + i\tilde r\ ,\quad
\chi = i\tilde \chi\ ,
\ee
which leads to
\bea
\label{FFF3}
ds^2 &=& d\tilde r^2 - {M \cosh^2 {2\tilde r \over l} \over \sqrt{M}\sinh {2\tilde r \over l}
+ 1}d\tilde \tau^2 \nn
&& + l^2 \left(  \sqrt{M} \sinh {2\tilde r \over l} + 1\right)
\left( d\tilde \chi^2 + \sinh^2 \tilde\chi \left(d\theta^2
+ \sin^2 \theta d\phi^2\right)\right) \ .
\eea
In the above metric (\ref{FFF3}), there is a time-like singularity at
$\sqrt{M} \sinh {2\tilde r \over l} = - 1$. If we put a boundary or a brane
source at
$\tilde r=R>0$ ($R$ can be a function of the ``time'' coordinate $\tilde \tau$ and
other coordinates, $\tilde\chi$, $\theta$, and $\phi$) and only consider the region
where $\tilde r>R$, then this procedure constrains the space-time  to
the non-singular region, only.
Since the
horizon is given by $\tilde r=0$ and it is  null, then for
the brane with a velocity  less than
 the speed of light, the brane does not cross the horizon and therefore
shields the space-time from the singularity.

Note that the  above suggested ways of obtaining cosmological solutions is
not limited to a narrow  class of
cosmological models  obtained from
 SAdS black holes but should also  apply to other AdS black holes, such as
 charged AdS  black
holes \cite{BehrndtCveticSabra} and  rotating charged AdS black
holes \cite{KlemmSabra}, and  they deserve further study.

We conclude this section by  deriving cosmological solutions from the
BPS bent domain wall solutions in a class of $N=2$
5-dimensional  gauged supergravity. Such  domain walls were found in
\cite{BC} (see also
\cite{Cardoso}). The world-volume of these  domain walls corresponds to the
$AdS_4$ space-time. Examples of such configurations have the property that
asymptotically ($y\to \infty$) the space-time is the boundary of $AdS_5$ while in the
interior ($y
\to y_0$)  there is a naked singularity.
One  such  a configuration  has the   metric of  the following form
\cite{BC}:
\be
\label{BC1}
ds^2 = \e^{-2y}\tilde g_{mn}dx^m dx^n + {4dy^2 \over \left(\e^{12y} - 1 \right)^2
 - 4\lambda^2 \e^{2y}}\ ,
\ee
and another is
\be
\label{BC2}
ds^2 = \e^{-2y}\tilde g_{mn}dx^m dx^n + {16dy^2 \over \e^{12y} - 16\lambda^2 \e^{2y}}\ .
\ee
Here $\tilde g_{mn}$ is the metric of 4d AdS, whose Ricci tensor $\tilde R_{mn}$
given by $\tilde g_{mn}$ satisfies $\tilde R_{mn} = - 3\lambda^2\tilde g_{mn}$.
There are several ways of expressing the  AdS metric, and  one of them is
\bea
\label{BC3}
\tilde g_{mn}(x) dx^m dx^n &=& - f(r) dt^2 + {1 \over f(r)} dr^2
+ r^2 \sum_{i,j=1}^2 \gamma_{ij} dx^i dx^j\ ,\nn
f(r)&\equiv& k + \lambda^2 r^2\ .
\eea
Here $k=0,\pm 1$. Consider $k=-1$ case. Then $\gamma_{ij}$ expresses
the hyperboloid with unit length parameter. When $\lambda^2 r^2 < 1$, by exchanging
$t$ and $r$, we obtain the following metric
\bea
\label{BC4}
\tilde g_{mn}(x) dx^m dx^n &=& - {1 \over \hat f(t)} dt^2 + \hat f(t) dr^2
+ t^2 \sum_{i,j=1}^2 \gamma_{ij} dx^i dx^j\ ,\nn
\hat f(t)&\equiv& 1 - \lambda^2 t^2\ ,
\eea
which may be reinterpreted as a cosmological model.

In the Euclidean signature, we may write the 4-dimensional AdS
(hyperboloid) as 
\be 
\label{BC5} 
\tilde g_{mn}(x) dx^m dx^n = dy^2
+ {1 \over \lambda^2}\cosh^2 \left(\lambda y\right) dH_3^2\ . \ee
Here $dH_3^2$ is the metric of the 3-dimensional hyperboloid with
a unit radius. Wick rotation $y=it$ yields a  cosmological model
with the following metric: \be \label{BC6} \tilde g_{mn}(x) dx^m
dx^n = -dt^2 + {1 \over \lambda^2}\cos^2 \left(\lambda t\right)
dH_3^2\ . 
\ee
Another comment is in order. Five dimensional backgrounds
with the asymptotically AdS space-time can be lifted on a 
five-sphere to 10-dimensions and   should correspond to Type IIB
supergravity backgrounds supported by a self-dual five-form field
strength $G_5$.  
The lift of SAdS BH to  D=10  gives for  $G_5\sim \epsilon_5$ where
$\epsilon_5$ is the volume form of the
$D=5$ metric:
\be
\label{Cvtc1}
G_5\sim \epsilon_5 = r^3 (\sin\chi)^2 (\sin\theta) dr\wedge dt \wedge
d\chi \wedge
d\theta \wedge d\phi\ .
\ee
In order for the cosmological solutions to have a consistent
Type IIB theory interpretation,   one has to check is that the
analytic continuation of $D=5$ metric
does not render the volume-form imaginary. 
We checked explicity the structure of
$G_5$   for all  the examples of analytic
continuations of SAdS BH solution and found that
the analytic
continuations in eqs.  
(\ref{AFN13}), (\ref{OT1}), (\ref{OT3}), (\ref{OT5}) (as well as the analytic continuation that 
leads to metric (\ref{FF3})) render  $G_5$ real.  Therefore all the
examples of analytic continuations yield consistent Type IIB backgrounds.

Since the bent domain wall solution involves the additional scalar
fields, that belong to the hypermultiplets of 
 a very specific  $D=5$, ${\cal N}=2$ gauged supergravity, the lift to
D=10 not known. However,  the part of $G_5$  that depends on
$\epsilon_5$ the analytic contuation yields  real  $\epsilon_5$.

In summary we presented a number of examples of cosmological AdS
space-times. Since these space-times  describe   (time-dependent)
AdS or Schwarzschild-AdS space-times, it  might be natural to
expect that some analog of AdS/CFT correspondence would be
applicable. However, due to time dependence and the related
effects, such as cosmological particle creation, the application
of the the AdS/CFT correspondence is not straightforward.
In the following sections we shall explore to what extent the AdS/CFT
correspondence can be applicable to this class of AdS  cosmological backgrounds.

\section{Holographic conformal anomaly
\label{s3}}

In this section we  investigate constraints on the time-dependent  AdS
space-times that reproduce the correct
 holographic trace anomaly.
The space-times with this feature  do not
reproduce the  components of the boundary stress-tensor, and thus the
standard AdS/CFT correspondence is not applicable.  Nevertheless, the fact that
the  trace  of the boundary stress-tensor can be reproduced  correctly is
an interesting feature of these backgrounds.

While we start with  a general consideration of  such space-times, we apply the
explicit calculation  of the trace anomaly  to the time-dependent solutions
obtained from the  so-called
 topological AdS black holes  ($k=-1$) \cite{CMZ}.
 We also comment on the fact that  the calculation of the conformal anomaly is
 applicable  also to the time-dependent backgrounds derived in the previous
 section. Subsequently, we also calculate  the thermodynamic energy for  the
 time-dependent  (topological) SAdS black hole background.

Using a general coordinate transformation,
one can start with  the following form of the metric:
\be
\label{HA1}
ds^2 = dy^2 + \e^{2y \over l}\sum_{m,n=0}^{d-1} \tilde g_{mn}(y,x) dx^m dx^n \ .
\ee
Defining a new coordinate $\rho$:
\be
\label{HA2}
\rho= \e^{- {2y \over l}}\ ,
\ee
the metric (\ref{HA1}) can be rewritten as
\be
\label{HA3}
ds^2 = {d\rho^2 \over 4\rho^2}+ \rho^{-1}\sum_{m,n=0}^{d-1} \tilde g_{mn}(y,x) dx^m dx^n \ .
\ee
The above metric is invariant under the following transformation with a constant parameter
$\lambda$ of the transformation:
\be
\label{HA3b}
\rho \to \lambda\rho \ ,\quad \tilde g_{mn} \to \lambda \tilde g_{mn}\ ,
\ee
which can be identified with a scale transformation for the metric $\tilde g_{mn}$.
First we consider the case that $\tilde g_{mn}$ and the Lagrangian density ${\cal L}$ can
be expanded as the power series on $\rho$:
\bea
\label{HA4}
\tilde g_{mn} (\rho,x) &=& \tilde g^{(0)}_{mn}(x) + \rho \tilde g^{(1)}_{mn}(x)
+ \rho^2 \tilde g^{(2)}_{mn}(x) + \cdots \ ,\nn
{\cal L} (\rho,x) &=& {\cal L}^{(0)}(x) + \rho {\cal L}^{(1)}(x)
+ \rho^2 {\cal L}^{(2)}(x) + \cdots \ .
\eea
Then the action has the following form:
\bea
\label{HA5}
S&=&\int d^{d+1}x\sqrt{-g}{\cal L} \nn
&=&{l \over 2}\int d^dx d\rho \rho^{-1 - {d \over 2}}
\left({\cal L}^{(0)}(x) + \rho {\cal L}^{(1)}(x)
+ \rho^2 {\cal L}^{(2)}(x) + \cdots \right)\ .
\eea
The $\rho$ integration of the terms containing ${\cal L}{(n)}$
with $n\leq {d \over 2}$ leads to a  divergence.
If $n<{d \over 2}$, the divergence can be subtracted  by adding
the counter-terms. The subtraction does not break the scale transformation in
(\ref{HA3b}). On the other hand, the subtraction of the term with $n={d \over 2}$
breaks the scale invariance in (\ref{HA3b}) and therefore gives the
holographic trace anomaly \cite{HS,CA}
\be
\label{HA6}
T = - l {\cal L}^{\left({d \over 2}\right)}\ .
\ee

Now let us assume that  the metric has the following, more general form than
that  in  (\ref{HA3}):
\be
\label{HA7}
ds^2 = {d\rho^2 \over 4\rho^2}+ \rho^\alpha \sum_{m,n=0}^{d-1}
\tilde g_{mn}(\rho,x) dx^m dx^n \ ,
\ee
One should note that the metric (\ref{HA7}) still has a invariance under a
kind
of scale transformaion, which is analogous to (\ref{HA3b}):
\be
\label{HA3c}
\rho \to \lambda^{1 \over \alpha} \rho \ ,\quad \tilde g_{mn} \to \lambda \tilde g_{mn}\ .
\ee
We now also assume $\tilde g_{mn} (\rho,x)$  (\ref{HA7}) can be expanded as a power series
in  $\rho$ as in (\ref{HA4}). Then the scalar curvature $R$ has the following form:
\be
\label{HA8}
R= - \alpha^2 d (d+1) + \rho^{-2\alpha}\tilde R\ .
\ee
Here $\tilde R$ is the scalar curvature constructed with $\tilde g_{mn}$.
Eq.(\ref{HA8}) implies that the behavior of the Lagrangian density ${\cal L}$
for small $\rho$ may be changed by the sign of $\alpha$. We now assume that the
Lagrangian density behaves as ${\cal L}\sim R$. Then in the  case of $\alpha<0$,
one can expand the Lagrangian as
\be
\label{HA9}
{\cal L} (\rho,x) = {\cal L}^{(0)}(x) + \rho {\cal L}^{(1)}(x)
+ \rho^2 {\cal L}^{(2)}(x) + \cdots
\ee
and the action has the following form:
\be
\label{HA10}
S={l \over 2}\int d^dx d\rho \rho^{-1 + {d\alpha \over 2}}
\left({\cal L}^{(0)}(x) + \rho {\cal L}^{(1)}(x)
+ \rho^2 {\cal L}^{(2)}(x) + \cdots \right)\ .
\ee
If ${d\alpha \over 2}$ is a non-positive integer, the term including
${\cal L}^{\left(-{d\alpha \over 2}\right)}$ diverges logarithmically.
If one subtracts this term, the scale invariance (\ref{HA3c}) breaks
down.
Then the term including ${\cal L}^{\left(-{d\alpha \over 2}\right)}$ may give
the holographic trace anomaly:
\be
\label{HA11}
T= -l {\cal L}^{\left(-{d\alpha \over 2}\right)}\ .
\ee
On the other hand, in the  case of $\alpha>0$, (\ref{HA8}) indicates the
following expansion of the Lagrangian:
\be
\label{HA12}
{\cal L} (\rho,x) = \rho^{-2\alpha}\left({\cal L}^{(0)}(x) + \rho {\cal L}^{(1)}(x)
+ \rho^2 {\cal L}^{(2)}(x) + \cdots \right)\ ,
\ee
and the action has the following form:
\be
\label{HA13}
S={l \over 2}\int d^dx d\rho \rho^{-1 + \left({d \over 2} -2\right)\alpha}
\left({\cal L}^{(0)}(x) + \rho {\cal L}^{(1)}(x)
+ \rho^2 {\cal L}^{(2)}(x) + \cdots \right)\ .
\ee
Then if $\left({d \over 2} -2\right)\alpha$ is non-positive integer, which requires
$d\leq 4$, the term including
${\cal L}^{\left(-{d\alpha \over 2}\right)}$ diverges logarithmically and may give
the holographic trace anomaly again:
\be
\label{HA14}
T= -l {\cal L}^{\left(\left({d \over 2} -2\right)\alpha\right)}\ .
\ee
For the special case: $d=4$, we have $\left({d \over 2}
-2\right)\alpha=0$. Then (\ref{HA6})
has the form
\be
\label{HA15}
T= -l {\cal L}^{\left(0\right)}\ .
\ee
In the case of the metric (\ref{dualmet}),
one may identify $\rho$ in (\ref{HA7}) as ${1 \over r^2}$ in
(\ref{dualmet}).
When $r$ is large ($\rho$ is small), the metric behaves as in
(\ref{SFN1}),
i.e., $(t,t)$-component of the metric is ${\cal O}\left(r^2\right)$ and
$(\xi^i,\xi^j)$-components is ${\cal O}\left(r^{-2}\right)$.
Then we should choose $\alpha=-1$ in (\ref{HA7}) from the $(t,t)$-component of the metric.
Then $(\xi^i,\xi^j)$-components of the metric corresponding to $\tilde g^{(0)}_{mn}$
(\ref{HA7}) vanish. Therefore the corresponding metric $\tilde g^{(0)}_{mn}$
(\ref{HA7}) cannot be defined
or becomes singular. On the other hand, the metric
(\ref{dualmet2}) for large $r$ behaves as
\be
\label{HA16}
ds^2 \to {l^2 \over r^2}dr^2 + {1 \over r^2}\left( - l^2 dt^2 + 2d\xi^2 \right)\ .
\ee
Defining
\be
\label{HA17}
\rho={1 \over r^2}\ ,
\ee
one gets
\be
\label{HA18}
ds^2 \to {l^2 \over 4\rho^2}d\rho^2 + \rho\left( - l^2 dt^2 + 2d\xi^2 \right)\ .
\ee
Since $\alpha=1>0$ and $d=4$, the quantity which we identify with the
holographic trace anomaly is
given by
(\ref{HA15}).

Let us now consider another metric:
\be
\label{DBH1}
ds^2 = dy^2 + l^2 \sinh^2 {y \over l}\sum_{m,n=0}^{d-1} \tilde g_{mn}(x) dx^m dx^n \ .
\ee
Redefining  the radial coordinate  $y$  as  in (\ref{HA2}),  and taking
 the limit  $\rho \to 0$, we have  (\ref{HA7}):
\footnote{
In \cite{HS}, the holographic trace anomaly has been obtained 
for the metric 
(\ref{DBH2}). We should note that the derivation does not depend on
whether the metric 
$\tilde g_{mn}$ depends on time or not.
}
\be
\label{DBH2}
ds^2 \to {d\rho^2 \over 4\rho^2}+ \rho^{-1} \sum_{m,n=0}^{d-1} \tilde g_{mn}(x) dx^m dx^n \ .
\ee
The $(d+1)$-dimensional vacuum Einstein equation with negative cosmological constant
requires that
\be
\label{DBH3}
\tilde R_{mn} = (d-1) \tilde g_{mn}\ .
\ee
We shall take as an explicit example the Schwarzschild-anti-deSitter black
hole which is
 a solution of (\ref{DBH2}), i.e.,
\bea
\label{DBH4}
\tilde g_{mn}(x) dx^m dx^n &=& - f(r) dt^2 + {1 \over f(r)} dr^2
+ r^2 \sum_{i,j=1}^{d-2} \gamma_{ij} dx^i dx^j\ ,\nn
f(r)&\equiv& {k \over d-3} - {\mu \over r^{d-3}} - r^2\ .
\eea
The metric $\gamma_{ij}$ satisfies the condition $R^{(\gamma)}_{ij}=k \gamma_{ij}$.
Here $R^{(\gamma)}_{ij}$ is the Ricci tensor given by $\gamma_{ij}$. The metric
has a curvature singularity at $r=0$. In fact,
\be
\label{DBH4b}
\tilde R_{klmn} \tilde R^{klmn} \sim
{(d-1)(d-2)^2(d-3) \over r^{2(d-1)}} \ .
\ee
In order for  space-time (\ref{DBH4}) to describe a black hole with
a horizon,
we should choose $k$ to be positive. When $k$ is not positive, the curvature
singularity in (\ref{DBH4}) becomes naked. However, the
interchange of the radial coordinate $r$ and the time coordinate $t$
yields a time-dependent (cosmological) solution, which  may be
referred to as  a topological
Schwarzschild-anti-deSitter
solution \cite{CMZ}:
\bea
\label{DBH5}
\tilde g_{mn}(x) dx^m dx^n &=& - {dt^2 \over {\tilde k \over d-3} + {\mu \over t^{d-3}} + t^2}
+ \left({\tilde k \over d-3} + {\mu \over t^{d-3}} + t^2\right) dr^2 \nn
&& + t^2 \sum_{i,j=1}^{d-2} \gamma_{ij} dx^i dx^j\ .
\eea
Here $\tilde k\equiv -k$. This $d$-dimensional space-time is embedded in
$d+1$-dimensional space-time as in (\ref{DBH1}) or (\ref{DBH2}).
In this case  the metric of the total $d+1$ dimensional space-time
is given by
\bea
\label{DBH1B}
ds^2 &=& = dy^2 + l^2 \sinh^2 {y \over l}\sum_{m,n=0}^{d-1} \tilde g_{mn}(x) dx^m dx^n \nn
&=& dy^2 + l^2 \sinh^2 {y \over l}\left(
- {dt^2 \over {\tilde k \over d-3} + {\mu \over t^{d-3}} + t^2}  \right. \nn
&& \left. + \left({\tilde k \over d-3} + {\mu \over t^{d-3}} + t^2\right) dr^2
+ t^2 \sum_{i,j=1}^{d-2} \gamma_{ij} dx^i dx^j\right)\ .
\eea
or
\bea
\label{DBH2B}
ds^2 &=& {d\rho^2 \over 4\rho^2}+ \rho^{-1} \left(
- {dt^2 \over {\tilde k \over d-3} + {\mu \over t^{d-3}} + t^2} \right.\nn
&& \left. + \left({\tilde k \over d-3} + {\mu \over t^{d-3}} + t^2\right) dr^2
+ t^2 \sum_{i,j=1}^{d-2} \gamma_{ij} dx^i dx^j\right)\ .
\eea
One can put a probe brane at $y=y_0$. The structure of the metric
(\ref{DBH1}) is similar to the case of deSitter
  brane in the AdS bulk space (see, for instance, Ref.\cite{Nunez}).
The stability of the brane may be achieved
by the quantum effects produced via the trace anomaly \cite{NO,NOZ,HHR}.

Using the standard procedure\cite{HS,CA}, the general
form of the holographic trace anomaly is:
\be
\label{DBH6}
T={l^3 \over 8\pi G}\left( {1 \over 8}\tilde R_{mn}\tilde R^{mn}-{1 \over 24}\tilde R^2
\right) \ .
\ee
Here $G$ is the $(d+1)$-dimensional Newton constant ($16\pi G=\kappa^2$).
For $d=4$ Eq.(\ref{DBH3}) gives
\be
\label{DBH7}
T= - {3 \over 16\pi G l}\ .
\ee
One should note that the obtained holographic trace anomaly exactly
coincides with
the trace anomaly calculated on the  CFT side.
Therefore even for  the time-dependent backgrounds of the type
(\ref{DBH1B}) or
(\ref{DBH2B}), the holographic trace anomaly is a constant.

We now consider the case that the space-time metric is given by
(\ref{FF3}).
Rewriting the metric with
\be
\label{TTT1}
\tilde r = {l \over 2}\ln \rho\ ,
\ee
as
\bea
\label{TTT2}
&& ds^2 = {l^2 d\rho^2 \over \rho^2} + {\left(1 + {4\mu \over l^2}\right)
\left({\rho - \rho^{-1} \over 2}\right)^2 \over \sqrt{1 + {4\mu \over l^2}}
{\rho + \rho^{-1} \over 2}- 1}d\tilde \phi^2 \nn
&& \quad + \left(  l^2\sqrt{1 + {4\mu \over l^2}} {\rho + \rho^{-1} \over 2} - l^2\right)
\left( d\chi^2 + \sin^2 \chi \left(d\theta^2 - \sin^2 \theta d\tilde \tau^2\right)\right) \ ,
\eea
in the limit of $\rho\to 0$, one gets
\bea
\label{TTT3}
ds^2 &=& {l^2 d\rho^2 \over 4\rho^2} + {1 \over \rho}\sum_{m,n=0}^3 \tilde g_{mn}(x)
dx^m dx^n\ ,\nn
\sum_{m,n=0}^3 \tilde g_{mn}(x)&=& \sqrt{1 + {4\mu \over l^2}}\left\{{1 \over 2}
d\tilde\phi^2 \right.\nn
&& \left. + l^2\left( d\chi^2 + \sin^2 \chi \left(d\theta^2 - \sin^2 \theta
d\tilde \tau^2\right)\right)\right\}\ .
\eea
Here $(x^m)=(\tilde\tau, \tilde \phi, \chi, \theta)$. The topology of the surface with
constant $\rho$ is $R\times dS^3$ and the radius of $dS^3$ is $l\sqrt{1 + {4\mu \over l^2}}$.
Since the metric (\ref{TTT3}) has a general form  (\ref{DBH2}), we can use the formula
(\ref{DBH6}). Since now
\be
\label{TTT4}
\tilde R_{\tilde\phi \tilde\phi}=0\ ,\quad
\tilde R_{ij} = {2 \over l^2 + 4\mu}\tilde g_{ij}\ ,
\ee
(Here $i,j$ corresponds to $(\tilde\tau, \chi, \theta)$), one finds the
conformal anomaly
vanishes
\be
\label{TTT5}
T=0\ ,
\ee
which is consistent with the field theory result.

Note that we can also apply  the calculation of the trace anomaly  to the
time-dependent backgrounds considered in the previous section. In particular,
for the time-dependent metric (\ref{t2}), which was obtained from the static one
(\ref{FF3}) by employing the analytic continuation
(\ref{FF2}),  the  metric can again be cast in the form  (\ref{TTT3}). Thus the
conformal anomaly can again be calculated  along the same lines, and it agrees
with the field theory result.
 Note
that the surface with constant $\rho$ corresponds to the space-like surface with
constant $\tilde \tau$,  and thus the  boundary field theory  lives on a
 space-like surface.

We now turn to the calculation of the thermodynamic  energy for the
time-dependent background (\ref{DBH1B}).
Note that  Eq.(\ref{DBH4b}) (with the exchanged  $r$ and $t$ coordinates)
 implies that
there is
a curvature singularity at $t=0$. As we are considering the case with
negative $k$, there is no horizon for
positive $\mu$ and the singularity is naked and    corresponds
to the big-bang (or possibly big-crunch) singularity. In the 5d metric (\ref{DBH1B}),
the singularity is $y$-dependent and the shape of the singularity is the line.

Since these metrics have naked singularities
 for positive $\mu$, it is not straightforward  to
define thermodynamical quantities.
If we choose, however, $\mu$ to be negative, there appears a cosmological
horizon.
For $d=4$, from Eq.(\ref{DBH5}), we find the horizon exists at $t=t_c$,
which
is defined by
\be
\label{DBH8}
0=\tilde k - {\tilde \mu \over t_c} + t_c^2\ ,\quad \tilde \mu \equiv -\mu
\ee
and one  can define the Hawking temperature $T_H$:
\be
\label{DBH9}
T_H \equiv {1 \over 4\pi}\left.{df(t) \over dr}\right|_{t=t_c}
= {1 \over 4\pi}\left({\tilde \mu \over t_c^2} + 2 t_c \right)=
{1 \over 4\pi}\left({\tilde k \over t_c} + 3 t_c\right)\ .
\ee
Assuming the entropy ${\cal S}$ is given by the horizon area, one finds
\be
\label{DBH10}
{\cal S}={r_c^2 V_2 \over 4G_4}\ ,\quad G_4\equiv {2G \over l}\ .
\ee
Here $V_2$ is the volume of the 2d manifold corresponding to $k=-\tilde k$ :
$V_2\equiv \int d^2 x \sqrt{\gamma}$.
When $V_2$ is infinite,  the density of the entropy may be considered.
Using the method of \cite{BBM}, the thermodynamical energy $E$ for $d=4$ is
nothing but $\tilde \mu$ \cite{CMZ}:
\be
\label{DBH11}
E={3\tilde \mu V_2 \over 16\pi G}=-{3\mu\ V_2 \over 16\pi G} .
\ee
In \cite{Cai}, the thermodynamical energy is related with the entropy
by modifying
the Cardy-Verlinde formula \cite{EV,cardy}. As the (topological)
black hole solution
exists at the constant $y$ surface, the entropy and  the energy
should correspond to those
on the CFT side.

We can also  calculate the analog of the surface energy momentum tensor
(\ref{sf13}) for
the
space-time
(\ref{DBH1B}) with $d=4$. Now
\be
\label{DBH12}
T_1^{mn} = {6 \over \kappa^2 l^3}{\cosh {y \over l} \over \sinh^3 {y \over l}}
\tilde g^{mn}\ ,\quad T_2^{mn}=\left(-{\eta_1 \over l^2 \sinh^2{y \over l}}
 - {6 \eta_2 \over l^4 \sinh^4 {y \over l}}\right)\tilde g^{mn}\ .
\ee
Choosing
\be
\label{DBH13}
\eta_1 = {6 \over \kappa^2 l}\ ,\quad \eta_2 = {l \over 2\kappa^2}\ ,
\ee
which is a standard choice in AdS/CFT correspondence \cite{BK},
the stress tensor in the large $y$ limit takes the form:
\be
\label{DBH14}
\hat T^{mn}=T_1^{mn} + T_2^{mn}\to - {48 \over 16\pi G l^3}\e^{-{6y \over l}}\tilde g^{mn}\ .
\ee
Then the trace of $\hat T^{mn}$ is given by
\be
\label{DBH15}
\hat T_m^{\ m}=g_{mn} \hat T^{mn} = l^2 \sinh^2 {y \over l}\tilde g_{mn} \hat T^{mn}
\to - {48 \over 16\pi G l^3}\e^{-{6y \over l}}\ ,
\ee
which is related  to  the holographic trace anomaly $T$ (\ref{DBH7}) by
\be
\label{DBH16}
T={\sqrt{-g_{(4)}} \over \sqrt{-\tilde g}}\hat T_m^{\ m}\to - {3l^3 \over 16\pi G}\ .
\ee
Here $g_{(4)mn}$ is the 4-dimensional part ($d=4$) of the metric (\ref{DBH2}):
$g_{(4)mn}=g_{mn}=l^2 \sinh^2{y \over l} \tilde g_{(4)}$.

The thermodynamical energy $\hat E$ can be evaluated from the bulk side by using the mass
formula (\ref{SFN14}). Now $q_t=\sqrt{-g_{tt}}=l\sinh {y \over l}
\sqrt{-\tilde g_{tt}}$ and other components of $q_\mu$ vanish. The
 expression for $\hat E$ follows:
\be
\label{DBH17}
\hat E=M={l^3 \over 16\pi G}{3 \over 4}\int d^3 x \sqrt{\gamma} t^2\ .
\ee
Here the integration $\int d^3 x\cdots $ is the integration with respect to
the spatial part of the 4d surface. We shall first consider the case with
$t>t_c$.
In order to avoid the orbifold singularity, when Wick-rotating the time
coordinate,  the
radial coordinate $r$ has to be periodic with the periodicity of the
inverse of the  Hawking temperature $T_H$ (\ref{DBH9}): $r\sim r + {1
\over T_H}$.
As a result
\be
\label{DBH18}
\hat E={l^3 \over 16\pi G}{3 \over 4}{V_2 \over T_H}t^2\ .
\ee
Note that the energy is time dependent.
When considering  the case with $t<t_c$, the role of  $t$ and $r$ is interchanged.
With  $r$ now as the time coordinate,
one can  define the integration
 $\int d^3 x \sqrt{\gamma} t^2$ as
$\int d^3 x \sqrt{\gamma} t^2=V_2\int_0^{t_c} dt t^2$. (Note,
there is a curvature singularity at $t=0$ and there might be a subtlety 
in the integartion close to $t=0$. However, since the
integrand is not singular at $t=0$, we  proceed by
assuming that we can integrate from $t=0$.) The energy is:
\be
\label{DBH19}
\hat E={l^3 \over 16\pi G}{t_c^3 V_2 \over 4}\ .
\ee
In the case  $\tilde k=-k=0$, Eq.(\ref{DBH8}) implies $\tilde \mu = t_c^3$ and
Eq.(\ref{DBH19}) can be rewritten as
\be
\label{DBH20}
\hat E={l^3 \over 16\pi G}{\tilde \mu V_2 \over 4}\ .
\ee
 One may costruct the entropy $\hat{\cal S}$ corresponding to the energy 
 $\hat E$  
using  thermodynamical formula $d\hat{\cal S}={d\hat E \over T_H}$. Since 
$\tilde k=-k=0$ 
\be
\label{AA1}
T_H = {3 \over 4\pi}t_c={3 \over 4\pi}\tilde\mu^{1 \over 3}\ .
\ee
Then we find
\be
\label{AA2}
d\hat{\cal S} = {l^3 V_2 \over 48}\tilde\mu^{-{1 \over 3}}d\tilde \mu\ 
\ee
and therefore
\be
\label{AA3}
\hat{\cal S}={l^3 V_2 \over 32G}\tilde\mu^{2 \over 3}+ {\cal S}_0
={l^3 V_2 \over 32G}t_c^2+ {\cal S}_0\ .
\ee
Here ${\cal S}_0$ is a constant of the integration. It is convenient to put
${\cal S}_0=0$. 
The free energy $\hat F$ is found to be
\be
\label{AA4}
\hat F = \hat E - T_H \hat{\cal S} = - {l^3 V_2 \tilde\mu \over 128\pi G}\ .
\ee
The obtained thermodynamical quantities satisfy the thermodynamical relations 
like the first law of the thermodynamics automatically by construction. 

The obtained expressions (\ref{DBH20}) and (\ref{AA3}) differ from (\ref{DBH11}) 
and (\ref{DBH10}) by a factor ${1 \over 12}$.
This discrepancy could be due to the fact that in the cosmological models
the definition of the thermodynamical quantities could be modified.
Another reason could be that definition of energy inside and
outside of $t_c$ is different. 
For deSitter-Schwarzschild black hole, there are two horizons, that is, 
the black hole horizon and 
the cosmological one. Since the areas of the two horizon are different, the entropies 
associated with the two horizons are different. In case of dS/CFT, the entropy and 
other thermodynamical quantities of the corresponding CFT agree with those associated 
with the cosmological horizon \cite{BBM}. This may be consistent with the conjecture 
that the CFT  exists on the space-like brane outside the cosmological horizon. 
Then possibly, the energy (\ref{DBH20})  corresponds to the energy of the CFT 
 living on the brane inside the horizon ($t<t_c$). 
It is  interesting that
there is similar phenomenon within standard  AdS/CFT
correspondence. Namely, in the calculation of  the  free energy
for  the 5d SAdS black hole  on the gravity side and on the dual
field theory side there is a  numerical  discrepancy. This
discrepancy  is attributed to the fact that the  gravity side the
calculation corresponds to  the result for a non-perturbative,
strongly coupled field theory, while the  calculation on the field
theory side is that of the perturbative (one-loop) CFT.

In conclusion, in this section we demonstrated for several
time-dependent  explicit examples that  some aspects of
 standard  AdS/CFT
correspondence  are  applicable. In particular,  we spelled
out the calculation
of the quantity which plays a role of the holographic anomaly on the
gravity side  and found agreement
with that on the CFT side.

\section{Scalar propagator
 \label{s4}}

In this section we shall calculate  the propagator of the scalar field
$\phi$ with mass $m$ in
the  time-dependent background
 (\ref{DBH1B}).   The aim is to  determine on the gravity side the  scaling
 dimensions of these fields and relate the results to the possible dual field
 theory interpretation.

For simplicity, the space with $k=-\tilde k=0$ is taken. We also
specialize to the region near the horizon. Note that the region far
from the horizon yields the results that are close to those in the
pure  AdS space-time.  For $k\neq 0$ case, the qualitative structure does not 
change and the same results maybe obtained.

As a starting point we redefine  the coordinate $t$ as
\be
\label{DBH21}
t=t_c + s\ ,
\ee
and  assume $0\leq s\ll t_c$. In this case  the Klein-Gordon equation has
the following form:
\bea
\label{DBH22}
0&=& \left(\Box - m^2\right)\phi \nn
&=& {1 \over \sinh^4{y \over l}}\partial_y \left(\sinh^4{y \over l}\partial_y \phi\right)
+ {1 \over l^2 \sinh^2{y \over l}}\left(-{3 \over t_c^3}\partial_s\left(s
\partial_s\phi\right) + {1 \over 3t_c^5 s}\partial_r^2 \phi\right) \nn
&& + {1 \over t_c^2l^2 \sinh^2{y \over l}}
\sum_{i=1,2}{\partial^2 \phi \over \partial {x^i}^2} -m^2\phi\ .
\eea
Introducing new coordinates $\xi$ and $\eta$ by
\be
\label{DBH23}
\xi=s^{1 \over 2}\cosh \left({3t_c \over 2}r\right)\ ,\quad
\eta=s^{1 \over 2}\sinh \left({3t_c \over 2}r\right)\ ,
\ee
one finds
\be
\label{DBH24}
-{3 \over t_c^3}\partial_s\left(s\partial_s\phi\right) + {1 \over 3t_c^5 s}\partial_r^2 \phi
= {3 \over 4t_c^3}\left( - {\partial^2 \over \partial \xi^2}
+ {\partial^2 \over \partial \eta^2}\right)\phi\ ,
\ee
which indicates that in this case the
 contribution associated with  coordinates $s$ and $r$  is that of the
flat space-time.
In the momentum space
one replaces   ${1 \over t_c^2}{\partial^2 \over \partial {x^i}^2}$ by
$-k^2$ and
${3 \over 4t_c^3}\left( - {\partial^2 \over \partial \xi^2}
+ {\partial^2 \over \partial \eta^2}\right)$ by $\omega^2$.
Then Eq.(\ref{DBH22}) can be rewritten as
\bea
\label{DBH25}
0 &=& \left(\Box_y - m^2\right)\phi \nn
\Box_y &\equiv& {1 \over \sinh^4{y \over l}}\partial_y \left(\sinh^4{y \over l}
\partial_y \phi\right)+ {\omega^2 - k^2 \over l^2 \sinh^2{y \over l}} \phi\ .
\eea
If we define the parameters $\mu$ and $\nu$ by
\be
\label{DBH26}
\mu^2={9 \over 4} - \omega^2 + k^2\ ,\quad
\nu\left(\nu + 1\right) = {15 \over 4} + l^2m^2 \ ,
\ee
or
\be
\label{DBH26b}
\mu=\sqrt{{9 \over 4} - \omega^2 + k^2} \ ,\quad
\nu = - {1 \over 2} + \sqrt{4 + l^2m^2}\ ,
\ee
the solutions of Eq.(\ref{DBH25}) are given by the associated Legendre
functions $P^\mu_\nu$ and $Q^\mu_\nu$:
\bea
\label{DBH27}
\phi(y) &=& \phi_P(y),\ \phi_Q(y) \ ,\nn
\phi_P(y) &\equiv& \left(\sinh{y \over l}\right)^{-{3 \over 2}}
P^\mu_\nu\left(\cosh{y \over l}\right)\ ,\nn
\phi_Q(y) &\equiv& \left(\sinh{y \over l}\right)^{-{3 \over 2}}
Q^\mu_\nu\left(\cosh{y \over l}\right)\ .
\eea
Since $P^\mu_\nu(x)\to \mbox{constant}$, $Q^\mu_\nu(x)\to \infty$ when $x\to 0$ and
$P^\mu_\nu(x)\to \infty$, $Q^\mu_\nu(x)\to 0$ when $x\to +\infty$, we may define
a propagator $G\left(y_1,y_2; \omega, k\right)$ by
\bea
\label{DBH28}
&& G\left(y_1,y_2; \omega, k\right) \nn
&& = C\left(\phi_P\left(y_1\right)\phi_Q\left(y_2\right)\theta\left(y_2 - y_1\right)
+ \phi_P\left(y_2\right)\phi_Q\left(y_1\right)\theta\left(y_1 - y_2\right)\right)\ .
\eea
Here $C$ is a normalization constant and $\theta(x)$ is a step function defined by
\be
\label{DBH29}
\theta(x)\equiv\left\{\begin{array}{ll}
1\quad & x>0 \\
0\quad & x<0 \\
\end{array}\right.\ .
\ee
In this case
\bea
\label{DBH30}
&& \Box_{y_1}G\left(y_1,y_2; \omega, k\right) \nn
&& = {C \over l\sinh^2 {y_1 \over l}}\left(-{P^\mu_\nu}'\left(\cosh{y_1 \over l}
\right)Q^\mu_\nu\left(\cosh{y_1 \over l}\right) \right. \nn
&& \quad \left. + P^\mu_\nu\left(\cosh{y_1 \over l}\right)
{Q^\mu_\nu}'\left(\cosh{y_1 \over l}\right)\right)\delta\left(y_1 - y_2\right) \nn
&& = -{C \over l\sinh^4 {y_1 \over l}}
{2^{2\mu}\e^{2\mu\pi i} \Gamma\left({\nu+\mu+1 \over 2}\right)
\Gamma\left({\nu + \mu \over 2} + 1\right) \over
\Gamma\left({\nu - \mu+1 \over 2}\right)\Gamma\left({\nu - \mu \over 2} + 1\right) }\ .
\eea
Here the following  formula for  the associated Legendre
functions was used:
\be
\label{DBH31}
-{d P^\mu_\nu\left(x\right) \over dx} Q^\mu_\nu\left(x\right)
+ P^\mu_\nu\left(x\right) {d Q^\mu_\nu\left(x\right)\over dx}
= {2^{2\mu}\e^{2\mu\pi i} \Gamma\left({\nu+\mu+1 \over 2}\right)
\Gamma\left({\nu + \mu \over 2} + 1\right) \over
\left(1 - x^2\right)\Gamma\left({\nu - \mu+1 \over 2}\right)
\Gamma\left({\nu - \mu \over 2} + 1\right) }\ .
\ee
Since $\sqrt{-g}\propto l^4\sinh^4{y \over l}$,
the normalization constant
$C$ can be chosen to be:
\be
\label{DBH32}
{1 \over C}= - {l^3 2^{2\mu}\e^{2\mu\pi i} \Gamma\left({\nu+\mu+1 \over 2}\right)
\Gamma\left({\nu + \mu \over 2} + 1\right) \over \Gamma\left({\nu - \mu+1 \over 2}\right)
\Gamma\left({\nu - \mu \over 2} + 1\right) }\ .
\ee

In the limit $x\to +\infty$, $P^\mu_\nu(x)$ and $Q^\mu_\nu(x)$ behave
as:
\be
\label{DBH32b}
P^\mu_\nu(x)\sim { \Gamma \left(\nu + {1 \over 2}\right) \left(2x\right)^\nu
\over
\sqrt{\pi}\Gamma\left(\nu - \mu + 1\right)}\ ,\quad
Q^\mu_\nu(x)\sim {\e^{\mu\pi i}\sqrt{\pi} \Gamma \left(\nu+\mu + 1\right) \over
\Gamma\left(\nu + {3 \over 2}\right)\left(2x\right)^{\nu+1}}\ .
\ee
For large $y$
\bea
\label{DBH33}
&& \phi_P\left(y\right) \sim
\e^{-\left({3 \over 2} - \nu\right){y \over l}}
=\e^{- \left( 2 - \sqrt{4 + l^2 m^2}\right){y \over l}}\ ,\nn
&& \phi_Q\left(y\right) \sim
\e^{-\left({5 \over 2} + \nu\right){y \over l}}
=\e^{- \left( 2 + \sqrt{4 + l^2 m^2}\right){y \over l}}\ .
\eea
Therefore the scalar field $\phi$ with  mass $m$ corresponds to the
conformal field
with the conformal weight $h$  given by
\be
\label{DBH34}
2h = 2 \pm \sqrt{4 + l^2m^2 }\ ,
\ee
Interesting, despite the fact that we started from  the cosmological AdS space, the
above result is in agreement with the standard  AdS$_5$/CFT$_4$ correspondence.
In the limit of $\rho\to 0$, 
the metric (\ref{DBH1B}) takes the form (\ref{TTT3}). It clear from the  
the arguments in section \ref{s3} that the  system has an 
invariance under the scale transformation (\ref{HA3b}) and thus
indicating that there is a dual  conformal field theory description. 

In the limit where $y_1=y_2=y\to \infty$,
$G\left(y_1,y_2; \omega, k\right)$ behaves as
\bea
\label{DBH37}
\lefteqn{G\left(y_1,y_2; \omega, k\right)\sim 8C\e^{-4{y \over l}}\e^{\mu\pi i}
{\Gamma\left(\nu + \mu +1\right) \over \left(\nu + {1 \over 2}\right)
\Gamma\left(\nu - \mu + 1\right)}} \nn
&=& - {\e^{-4{y \over l}}\Gamma\left({\nu - \mu+1 \over 2}\right)
\Gamma\left({\nu - \mu \over 2} + 1\right) \over
l^3 2^{2\mu}\e^{\mu\pi i} \Gamma\left({\nu+\mu+1 \over 2}\right)
\Gamma\left({\nu + \mu \over 2} + 1\right) }
{\Gamma\left(\nu + \mu +1\right) \over \left(\nu + {1 \over 2}\right)
\Gamma\left(\nu - \mu + 1\right)} \ ,
\eea
which has poles at
\be
\label{DBH38}
\nu - \mu +1 = -2N\ ,\quad N=0,1,2,3,\cdots \ .
\ee

 Employing  (\ref{DBH26}) or (\ref{DBH27}),
we obtain the following relationship:
 \be \label{DBH39} \omega^2 -
k^2 = m_\omega^2 = {9 \over 4} + \left\{{1 \over 2} + 2N + \sqrt{4
+ l^2m^2} \right\}\ . \ee
Eq.(\ref{DBH39}) gives an effective mass in the 4d
space-time  associated with  $(t,r,x^i)$ coordinates.
Using (\ref{DBH23}),  the time evolution of the scalar field $\phi$ is of the form
\be
\label{DBH39b}
\phi \propto \e^{i\left(-\omega_\xi \xi + \omega_\eta \eta\right)}
= \e^{-i \omega s^{1 \over 2}\cosh\left( {3t_c \over 2}r - \theta_0\right)}\ .
\ee
Here the components of $\omega$ are written as
\be
\label{DBH39c}
\omega_\xi = \omega \cosh \theta_0\ ,\quad \omega_\eta = \omega \sinh \theta_0\ .
\ee

A massless scalar field $\phi$ depends only on the  coordinate $t$ or $s$.
Then Eq.(\ref{DBH22}) reduces as
\be
\label{DBH40}
0=\partial_s\left(s\partial_s\phi\right) \ ,
\ee
whose solution is given by
\be
\label{DBH41}
\phi=\alpha_0\ln {s \over s_0}\ .
\ee
Here $\alpha_0$ and $s_0$ are constants of the integration.
The  solution (\ref{DBH41}) diverges at the horizon $s=0$.
In fact, Eq.(\ref{DBH41}) has the form of the scalar propagator in
2d  space-time, however with the non-trivial time dependence.

The above results indicate again that the scalar propogator in
the time-dependent AdS backgrounds has the structure that indicates the scaling
dimensions of the  fields in agreement with the standard AdS/CFT correspondence.

\section{Particle creation \label{s5}}

Time-dependent backgrounds often imply  multiple
vacua.  In such cases one  vacuum should be an excited state
of  another
vacuum. In particular  when the out-vacuum, corresponding to an out-state,
is different from the in-vacuum, corresponding to an in-state,
the particle creation effects occur. 
\footnote{There are, of course, ambiguities of the choice of the vacua. 
In case of the cosmological spacetime, it would be natural to choose the vacua 
corresponding to the cosmological time or the (monotonically increasing) function 
of the cosmological time. }
For example,  a real scalar
field may be expanded by  a set of solutions $\left\{
u_k^{\rm in}\right\}$ or $\left\{ u_k^{\rm out}\right\}$  of
the Klein-Gordon equation as
\bea
\label{OP1}
\phi &=& \sum_k \left\{ a_k^{\rm in} u_k^{\rm in} + a_k^{{\rm in}*}
u_k^{{\rm in}*} \right\}\nn
&=& \sum_k \left\{ a_k^{\rm out} u_k^{\rm out} + a_k^{{\rm out}*} u_k^{{\rm out}*}\right\}
\eea
 Here  annihilation operators are denoted by
$a_k^{\rm in,out}$ and  creation operators by $a_k^{{\rm
in,out}*}$.
If  two sets of  solutions are related by
\be
\label{OP2}
u_k^{\rm in} = \alpha_k u_k^{\rm out} + \beta_k u_k^{{\rm out}*}\ ,
\ee
we have
\be
\label{OP3}
a_k^{\rm out} = \alpha_k a_k^{\rm in} + \beta_k^* a_k^{{\rm in}*}\ .
\ee
Since the number operator corresponding to $k$ in the out-mode is
$N_k^{\rm out}=a_k^{{\rm out}*}a_k^{\rm out}$, the number of particles of this
mode in  the in-vacuum is given by $\left|\beta_k\right|^2$.

One may consider the two-point functions of two $\phi$,
as in \cite{BLN}:
\be
\label{CF1}
_{\rm in}\langle 0| \phi(1) \phi(2) |0\rangle_{\rm in}
= \sum_k u_k^{{\rm in}}(1) u_k^{{\rm in}*}(2)\ .
\ee
Here we denote the coordinates of the respective fields  by 1 or 2. On
the other hand,
\bea
\label{CF2}
&& _{\rm out}\langle 0| \phi(1) \phi(2) |0\rangle_{\rm in}
= \sum_k {u_k^{{\rm out}}(1) u_k^{{\rm in}*}(2) \over \alpha_k}
{_{\rm out}\langle} 0|0\rangle_{\rm in} \nn
&& =\left\{_{\rm in}\langle 0| \phi(1) \phi(2) |0\rangle_{\rm in}
+ \sum_k {\beta_k \over \alpha_k}u_k^{{\rm out}}(1) u_k^{{\rm in}*}(2)\right\}
{_{\rm out}\langle} 0|0\rangle_{\rm in}\ .
\eea
For the propagator,  the same relation follows:
\bea
\label{CF3}
 iG_{F{\rm out}\leftrightarrow {\rm in}}(1,2)
&=& iG_{F{\rm in}\leftrightarrow {\rm in}}(1,2)
+ \sum_k {\beta_k \over \alpha_k}u_k^{{\rm out}}(1) u_k^{{\rm in}*}(2)
{_{\rm out}\langle} 0|0\rangle_{\rm in} \nn
iG_{F{\rm out}\leftrightarrow {\rm in}}(1,2)&\equiv&
{_{\rm out}}\langle 0| T\phi(1) \phi(2) |0\rangle_{\rm in}\ ,\nn
iG_{F{\rm in}\leftrightarrow {\rm in}}(1,2)&\equiv&
{_{\rm in}\langle} 0| T\phi(1) \phi(2) |0\rangle_{\rm in}\ .
\eea
Here $T$ is  the time ordering operator.

Recent interest in the AdS/CFT correspondence in
time-dependent
backgrounds \cite{NOcosm, adstime}, motivated the study  \cite{NOjcap,AC} of
particle creation  for  the 4d deSitter brane, which can
be regarded  as an inflationary Universe, embedded in
the 5d AdS bulk. Within the AdS/CFT
correspondence, the partition function in the bulk gravity corresponds
to the generating function of the dual CFT. The CFT generating
function for the operator ${\cal O}$ between the in- and the out-vacuum is
given by the bulk partition function $Z_{{\rm in}\to{\rm out}} \left(
\phi_0 \right)$, which is calculated   by imposing proper
initial and final  boundary  conditions for the field in the bulk
\be
\label{OP4}
_{\rm out}\langle 0|T\e^{i\int_{\rm boundary} \phi_0 {\cal O}}| 0 \rangle_{\rm in}
=Z_{{\rm in}\to{\rm out}} \left( \phi_0 \right)\ .
\ee
Let us assume that the metric of the bulk space-time has the following
form:
\be
\label{OP5}
ds^2={1 \over z^2}\left(dz^2 + \tilde g_{mn}dx^m dx^n\right)\
\ee
and assume that there is a boundary at $z=0$. Then the bulk-boundary
propagator $G_B\left(x^m,z;{x^m}'\right)$, which connects the boundary value $\phi_0$
of the scalar field $\phi$, as
\be
\label{OP6}
\phi\left(x^m,z\right)=\int_{\rm boundary}dx' \sqrt{-\tilde g}
G_B\left(x^m,z;{x^m}'\right)\phi_0\left({x^m}'\right)\ ,
\ee
is given by the bulk propagator $G_F$ \cite{giddings}
\be
\label{OP7}
G_B\left(x^m,z;{x^m}'\right)\phi_0\left({x^m}'\right)
=2\hat \nu \lim_{z'\to 0}{z'}^{-2h_+ }G_F\left(x^m,z;{x^m}',z'\right)\phi_0\left({x^m}'\right)\ .
\ee
Here $h_\pm =1 \pm {\nu \over 2}$, $\hat \nu=\sqrt{4 + m^2 l^2}$ ($m$ is the mass of the
scalar field). In \cite{BLN}, by combining (\ref{CF3}), (\ref{OP4}), and (\ref{OP7}),
the following expression is obtained
\bea
\label{OP8}
&&{_{\rm in}\langle} 0| \phi(1) \phi(2) |0\rangle_{\rm in} \nn
&&={{_{\rm out}\langle} 0| \phi(1) \phi(2) |0\rangle_{\rm in}
\over {_{\rm out}\langle} 0|0\rangle_{\rm in}} \nn
&& \quad + 2i\hat \nu \lim_{z\to 0,\ \epsilon\to 0}\epsilon^{1 - 2h_+}\left.
{\partial \over \partial z'}
\left\{{z}^{-2h_+} u_k^*\left(x^m,z\right) u_k^*\left({x^m}',z'\right)\right\}
\right|_{z'=\epsilon}\ .
\eea
The second term is responsible for particle creation effect
when applying AdS/CFT correspondence to the time-dependent
backgrounds.

In this section, we investigate the particle production for the scalar
field $\phi$ satisfying the Klein-Gordon equation (\ref{DBH22})
in the space-time  background (\ref{DBH1B}).
Eqs.(\ref{DBH22}) and (\ref{DBH24}) indicate that $\phi$ satisfies
\be
\label{PP1}
-{3 \over t_c^3}\partial_s\left(s
\partial_s\phi\right) + {1 \over 3t_c^5 s}\partial_r^2 \phi
={3 \over 4t_c^3}\left( - {\partial^2 \over \partial \xi^2}
+ {\partial^2 \over \partial \eta^2}\right)\phi= \omega^2\phi\ .
\ee
In the coordinates $\xi$ and $\eta$, the 2d space-time is
flat. When one regards  $\xi$ as a time coordinate, there is no particle
creation.  However, the natural time coordinate should be $s$,
which is related to the  global time coordinate $t$. As
the coordinate
system given by $s$ and $r$ is
not inertial one,
the particle creation takes place.

For  $\phi\propto \e^{-ipr}$, where $p$ is a constant corresponding to
the momentum conjugate to $r$, Eq.(\ref{PP1}) reduces as
\be
\label{PP2}
0={3 \over t_c^3}\partial_s\left(s
\partial_s\phi\right) + {p^2 \over 3t_c^5 s} \phi
+ \omega^2 \phi\ .
\ee
With the redefinition of  $s$ in terms of $
\sigma$:
\be
\label{PP3}
s={\sigma^2 \over \omega^2 }\ ,
\ee
one obtains  the Bessel differential equation:
\be
\label{PP4}
0={\partial^2 \phi \over \partial \sigma^2} + {1 \over \sigma}{\partial \phi
\over \partial \sigma} + \left({\tilde \nu^2 \over \sigma^2} + 1 \right)\phi\ ,
\quad \tilde \nu^2 = {4p^2 \over 9t_c^2}\ ,
\ee
whose solution can be obtained via the Hankel functions $H_{i\tilde \nu}^{(1)}(\sigma)$
and $H_{i\tilde \nu}^{(2)}(\sigma)$. Since
$0\leq s <+\infty$, then
 $0\leq \sigma <
+\infty$.
In the following, however, we analytically continue $\sigma$ into the region where
$\sigma$ is negative, i.e. $-\infty<\sigma<\infty$, and regard  the
limit
$\sigma\to + \infty$  to  correspond to the out-state and  $\sigma\to -\infty$
to the in-state.

For  positive and large $\sigma$,
the Hankel functions $H_{i\tilde \nu}^{(1)}(\sigma)$ and $H_{i\tilde \nu}^{(2)}(\sigma)$
behave as
\be
\label{PP5}
H_{i\tilde \nu}^{(1)}(\sigma) \sim \sqrt{{2 \over \pi\sigma}}\e^{i\left(\sigma
 - {\left(2i \tilde \nu +1\right) \over 4}\right)}\ ,\quad
H_{i\tilde \nu}^{(2)}(\sigma)\sim \sqrt{{2 \over \pi\sigma}}\e^{-i\left(\sigma
 - {\left(2i \tilde \nu +1\right) \over 4}\right)}\ .
\ee
In this limit   $H_{i\tilde \nu}^{(1)}(\sigma)$ correspond to the
 negative frequency modes of the out-vacuum and
$H_{i\tilde \nu}^{(2)}(\sigma)$ to  positive frequency ones.
Since the Bessel differential equation (\ref{PP4}) is invariant
under  analytical
continuation: $\sigma \to -\sigma$, then $H_{i\tilde \nu}^{(1)}(-\sigma)$
and
$H_{i\tilde \nu}^{(2)}(-\sigma)$ are also solutions. Eq.(\ref{PP5})
then implies that
$H_{i\tilde \nu}^{(1)}(-\sigma)$ correspond to   positive frequency
modes of the in-vacuum  and
$H_{i\tilde \nu}^{(2)}(\sigma)$ to  negative frequency ones.
Therefore
\bea
\label{PP6}
\phi &=& \sum_{\tilde\nu}\left( b^{{\rm out}*}_\nu H_{i\tilde \nu}^{(1)}(\sigma)
+ a^{{\rm out}}_\nu H_{i\tilde \nu}^{(2)}(\sigma)\right) \nn
&=& \sum_{\tilde\nu}\left( b^{{\rm in}*}_\nu H_{i\tilde \nu}^{(2)}(-\sigma)
+ a^{{\rm in}}_\nu H_{i\tilde \nu}^{(1)}(-\sigma)\right)\ .
\eea
The out-vacuum $|0\rangle_{\rm out}$ and in-vacuum $|0\rangle_{\rm in}$
can be defined by
\be
\label{PP7b}
a^{{\rm out}}_\nu |0\rangle_{\rm out}= b^{{\rm out}}_\nu |0\rangle_{\rm out}=0\ ,\quad
a^{{\rm in}}_\nu |0\rangle_{\rm out}= b^{{\rm in}}_\nu |0\rangle_{\rm in}=0\ .
\ee
The analytic continuation $\sigma \to \e^{i\theta}\sigma$ and  taking
$\theta=\{0,1\}$  implies that for
 $\sigma \to -\sigma$ one gets
\bea
\label{PP7}
H_{i\tilde \nu}^{(1)}(-\sigma)&=& - \e^{\tilde\nu\pi}H_{i\tilde \nu}^{(2)}(\sigma) \nn
H_{i\tilde \nu}^{(2)}(-\sigma)&=&
2\cosh\left(\tilde\nu \pi\right)H_{i\tilde \nu}^{(2)}(\sigma)
 + \e^{-\tilde\nu\pi}H_{i\tilde \nu}^{(1)}(\sigma)
\eea
Then the coefficients of the Bogolubov transformation between the
in- and out-vacuum are of the form:
\be
\label{PP8}
b^{{\rm out}*}_\nu = \e^{-\tilde\nu\pi}b^{{\rm in}*}_\nu  \ ,\quad
a^{{\rm out}}_\nu = 2\cosh\left(\tilde\nu \pi\right) b^{{\rm in}*}_\nu
 - \e^{\tilde\nu\pi}a^{{\rm in}}_\nu \ ,
\ee
or
\be
\label{PP8b}
b^{{\rm in}*}_\nu= \e^{\tilde\nu\pi}b^{{\rm out}*}_\nu   \ ,\quad
a^{{\rm in}}_\nu = - \e^{-\tilde\nu\pi}a^{{\rm out}}_\nu
+ 2\cosh\left(\tilde\nu \pi\right)b^{{\rm out}*}_\nu
\ee
The number $N_{\tilde \nu}$ of created out-particles corresponding to $\tilde \nu$ in
in-vacuum is given by
\be
\label{PP9}
N_{\tilde \nu} = _{\rm in}\langle 0 |\left\{ a^{{\rm out}*}_\nu a^{{\rm out}}_\nu
+ b^{{\rm out}*}_\nu b^{{\rm out}}_\nu \right\}|0\rangle_{\rm in}
= 4 \cosh^2\left(\tilde\nu \pi\right)\ ,
\ee
which becomes large for large $\tilde\nu$. From Eq.(\ref{PP4}),
large $\tilde \nu$  implies large momentum  $p$.

 Note that this calculation is done by
 analytically continue $\sigma$  into unphysical region
 with negative $\sigma$, while
 the   original region  for $\sigma$ was: $0\leq \sigma < +\infty$.
The restriction to the  physical region $0\leq \sigma < +\infty$,
  would imply that
$N_{\tilde \nu}$  is now smaller.
While we were unable to obtain  analytic results for $N_{\tilde\nu}$ in the
physical region
 $0\leq \sigma < +\infty$, let us make the following comment.
Since the Hankel functions $H_{i\tilde \nu}^{(1,2)}(\sigma)$ behave as
$\sigma^{\pm i\nu}$ when $\sigma$ is small,  one could  define a new
time coordinate $\tilde\sigma$ by $\tilde \sigma = \ln \sigma + \sigma$.
The behavior of the Hankel functions
when $\sigma$ or $\tilde \sigma$ is large is not changed: $H_{i\tilde \nu}^{(1,2)}(\sigma)
\sim H_{i\tilde \nu}^{(1,2)}(\tilde\sigma)$ but when $\sigma$ is small, we find
\bea
\label{0inf1}
\sinh \left(\tilde\nu \pi\right) H_{i\tilde \nu}^{(1)}(\tilde \sigma)
&\sim& -2^{i\tilde \nu \pi}\e^{-i\tilde \nu \tilde \sigma}
+ 2^{-i\tilde\nu}\e^{\tilde\nu}\e^{i\tilde \nu \tilde \sigma} \nn
\sinh \left(\tilde\nu \pi\right) H_{i\tilde \nu}^{(2)}(\tilde \sigma)
&\sim& 2^{i\tilde \nu}\e^{-i\tilde \nu \tilde \sigma}
+ 2^{-i\tilde\nu}\e^{-\tilde\nu\pi }\e^{i\tilde \nu \tilde \sigma} \ .
\eea
The part behaving as $\e^{-i\tilde \nu \tilde \sigma}$
($\e^{i\tilde \nu \tilde \sigma}$) can  be identified with the positive
(negative)
frequency part. Then the Bogolubov coefficient $\beta$ in (\ref{OP2})
behaves exponentially $\beta\sim \e^{\tilde\nu \pi}$ again.
In any case, since  for the region
$-\infty<\sigma<\infty$, there is a symmetry $\sigma\to -\sigma$, we expect,
that  in the interval $0<\sigma<\infty$,
$N_{\tilde \nu}$ is reduced, naively,  by a factor of two relative to
(\ref{PP9}).

At first sight the result that $N_{\tilde\nu}$ grows exponentially
with $\tilde \nu$   is puzzling, since for example in the
deSitter space, creation of particles with high momenta is
exponentially suppressed \cite{mottola}.
The resolution to this puzzle lies in the fact that the
calculation was performed  originally (see Eq.(\ref{DBH21})) only in 
the region $|s|\ll t_c$  and it turns out the values of
$\tilde\nu$ are bounded from below to be $\le {\cal O}(1)$. This
can be understood in the following way. Note that $t_c$ plays a
role of a cutoff scale.
When $t\sim t_c$, the  4d slice of the metric (\ref{DBH5}) with
constant $y$
has the following form:
\be
\label{PPP1}
\tilde g_{mn}(x) dx^m dx^n \sim - {ds^2 \over 4\pi T_H s} + 4\pi T_H s dr^2 \nn
 + t_c^2 \sum_{i,j=1}^{d-2} \gamma_{ij} dx^i dx^j\ .
\ee
The form of the metric (\ref{PPP1}) implies  that in this limit
  the scale  factor associated with $s$ is large, and for the
the  radial  direction $r$ the scale factor is small (inversely proportional to
the scale factor of $s$).  Thus, the  large scale of  $s$ (or $t$)   corresponds to the small
(inverse) scale of  $r$.   Since  $p$ in (\ref{PP2}) is a
momentum, dual to the radial direction $r$,
it would  scale  as $t$,  which is consistent with the
dimensionless expression of $\tilde \nu$ in (\ref{PP4}). The
expression of $\tilde \nu$ in (\ref{PP4}) also implies  that
$\tilde \nu$ cannot be large. Thus, the calculation of
$N_{\tilde \nu}$  (\ref{PP9}) is   valid only for   $\tilde \nu
\le {\cal O}(1)$. On the other hand
 for large $\tilde\nu$, one should consider the region $t\gg t_c$, where
 the metric of the slice with constant $y$
is pure deSitter. In this case,  the particle production for large
$\tilde\nu$ should be suppressed exponentially \cite{mottola},
which is consistent with the result in \cite{BLN}. Thus, the total
number of  the produced particles should become finite. While the
above considerations shed light on  the qualitative picture of
particle production for this background, it would be interesting
to obtain the quantitative result for $N_{\nu}$ in the full range
of $t$.

 Let us now  consider  the last term in (\ref{OP8}).
>From the metric (\ref{DBH1B}), one may identify $z=
l^{-1}\sinh^{-1}{y \over l}\sim 2 \e^{-{y \over l}}$. We also
assume the conformal weight $h_+$ in (\ref{OP8}) as $h$ in
(\ref{DBH34}) with $(+)$-sign and $\hat \nu$ as $\sqrt{4+l^2
m^2}$. It follows  that $u_k^*$ is given by combining
(\ref{DBH27}) and (\ref{PP6}), etc., as: \be \label{PP10} u_k^*
\sim \sqrt{C} \e^{ipr}H^{(2)}_{i\tilde \nu}(-\sigma)
\e^{it_c\sum_{i=1,2}k_i x^i} \left(\sinh{y \over l}\right)^{-{3
\over 2}} Q^\mu_\nu\left(\cosh{y \over l}\right)\ . \ee Here $C$
is a normalization constant  and $x^i$'s ($i=1,2$) are coordinates
defined in (\ref{DBH1B}). It follows from (\ref{DBH33}), that
$\phi_Q\left(y\right)$ corresponds to the conformal field with the
weight $h$ with $(+)$\, signs. In this case  the last term in
(\ref{OP8}) has the following form: \bea \label{PP11}
&&2i\sqrt{4+l^2 m^2}C\e^{ip(r+r')}H^{(2)}_{i\tilde \nu}(-\sigma)
H^{(2)}_{i\tilde \nu}(-\sigma') \e^{it_c\sum_{i=1,2}k_i \left(x^i
+ {x^i}'\right)} \nn && \times \left({5 \over 2} + \sqrt{4+l^2
m^2}\right) 2^{-2 -2 \sqrt{4+l^2 m^2}} \nn && \times {\pi
\e^{2\mu\pi i} \Gamma\left(\nu+\mu+1\right)^2 \over
\Gamma\left(\nu+{3 \over 2}\right)^2}\ . \eea Here $\nu$ and $\mu$
are defined in (\ref{DBH26b}). $C$ should also depend on $\nu$ and
$\mu$ as in (\ref{DBH32}). If $C$ is given by $|C|$ in
(\ref{DBH32}), Eq.(\ref{PP11}) is rewritten as \bea \label{PP12}
&&2i\sqrt{4+l^2 m^2}{\Gamma\left({\nu - \mu+1 \over 2}\right)
\Gamma\left({\nu - \mu \over 2} + 1\right) \over l^3
2^{2\mu}\e^{2\mu\pi i} \Gamma\left({\nu+\mu+1 \over 2}\right)
\Gamma\left({\nu + \mu \over 2} + 1\right) } \nn && \times
\e^{ip(r+r')}H^{(2)}_{i\tilde \nu}(-\sigma) H^{(2)}_{i\tilde
\nu}(-\sigma') \e^{it_c\sum_{i=1,2}k_i \left(x^i + {x^i}'\right)}
\nn && \times \left({5 \over 2} + \sqrt{4+l^2 m^2}\right) 2^{-2 -2
\sqrt{4+l^2 m^2}} \nn && \times {\pi \e^{2\mu\pi i}
\Gamma\left(\nu+\mu+1\right)^2 \over \Gamma\left(\nu+{3 \over
2}\right)^2}\ . \eea In the above expressions (\ref{PP11}) and
(\ref{PP12}), $y$-dependence is cancelled. In order to obtain the
final expression, we have  to integrate the obtained expressions
with respect to $p$ (or $\tilde \nu$), $\omega$, and  $k$.
However,  we were unable to obtain the  explicit integration
results. One may be able to determine  dominant regions of the
integral and obtain an approximate expression. In any case,
 (\ref{PP11}) or (\ref{PP12}) imply that there will be
specific modifications on the
 CFT side, due to particle creation.

The geometry of the boundary where the CFT  is given by
(\ref{DBH5}), with $\tilde k=0$ and $\mu=-\tilde \mu$. When
$t=t_c$, the coordinate $r$ is degenerate and the spatial region
becomes 2-dimensional flat space given by the coordinates $x^i$.
If we compactify $x^i$, the spatial topology can be a two-torus.
When $t$ is large, the space-time geometry becomes 3-dimensional
deSitter space. If we compactify the coordinates $r$ and $x^i$,
the geometry of the spatial part is S$^1\times$T$^2 \sim $T$^3$.
Here S$^1$ corresponds to $r$ and T$^2$ to $x^i$. The radius of
S$^1$ is proportional to the inverse of time $t$ and that of T$^2$
to time $t$ itself.

Creation of particles gives an extra contribution to the energy-momentum
tensor
and it will shift the energy of the vacuum. Then it is natural to expect that the
surface stress-tensor (\ref{DBH12}) will also be  modified by the
contribution from
the created particles, which might resolve the discrepancy between the expressions of
energy in (\ref{DBH11}) and (\ref{DBH20}). It is interesting, however,
that the trace of the
stress-tensor (conformal anomaly) will be the same for  any (in- or out-)
state
which was actually demonstrated in section \ref{s3}.

\section{Wilson loop \label{s6}}

In this section the Wilson loop \cite{rey} is found in the
background (\ref{DBH1B}) with $\tilde k=0$. The formal procedure
treating the Wilson loop is a standard one \footnote{Let us stress
again, that the interpretation of  this quantity, calculated on
the gravity side, as a Wilson loop is valid only if the AdS/CFT
correspondence is applicable.}. First one considers a loop on the
boundary ($y\to \infty$ in (\ref{DBH1B})) of the bulk space-time
and finds a configuration of the string, whose boundary is the
loop, so that the Nambu-Goto action (\ref{SFN20}) has its minimum.
For the case that the space-time is static, we may take a
rectangular loop, whose length is $T$ and $L$ in the (Euclidean)
time direction  and in the spatial direction, respectively. If
$T\gg L$, the potential energy between a quark and an anti-quark
is given by dividing the value of the action, evaluated for the
string configuration, by $T$. Since the space-time under
discussion is now time-dependent, the metric inside the Nambu-Goto
action (\ref{SFN20}) becomes time-dependent as well.
 Moreover, we cannot  consider  only rectangular loops.
In principle, one should  consider a loop of an arbitrary shape,
which is the  boundary of the string.  After that
 one divides    the
string by the small time interval. In the time interval, the time
distance between quark and anti-quark can be defined. Furthermore,
in this time interval one may often neglect the time dependence
using an adiabatic approximation. By such a procedure, the
potential between a quark and an anti-quark can be  defined.

We start with  the near-horizon  region ($0\leq s\ll t_c$) for
(\ref{DBH21}).
Then the metric (\ref{DBH1B}) has a following form:
\be
\label{W1}
ds^2 \sim dy^2 + l^2 \sinh^2 {y \over l}\left( - {ds^2 \over 3t_c s} + 3t_c s dr^2
+ t_c^2 \sum_{i,j=1,2} \left(dx^i\right)^2\right)\ .
\ee
We also employ Wick rotation  and the following  redefinition of  the
coordinates
\be
\label{W2}
s=-{3t_c \over 4}u^2\ ,\quad r={2i \over 3t_c}v, \quad x^i = {1 \over t_c}\hat x^i\ .
\ee
Since the region near the boundary ($y\to +\infty$) is taken, it is assumed $y$
is large. The corresponding metric is:
\be
\label{W3}
ds^2 = dy^2 + {l^2 \over 4}\e^{2y \over l}\left( du^2 + u^2 dv^2
+ \sum_{i,j=1,2} \left(d\hat x^i\right)^2\right)\ .
\ee
We  consider a short time interval given by $u_0\leq u \leq u_0 + \Delta
u$
($0<\Delta u \ll u_0$) and thus  $u$ is close to being  constant.
In this case  the Nambu-Goto string action (\ref{SFN20}) can be
evaluated, by choosing  $\tau=u$ and $\sigma=v$ in (\ref{SFN20}). As $u$
is almost constant,
 the configuration   does not depend on $u$ in
the interval $u_0\leq u \leq u_0 + \Delta u$, as $y=y(v)$ and $\hat x^i=0$ is taken.
Then the Nambu-Goto action has the following form:
\be
\label{W4}
S_N = {\Delta u l \over 2}\int dv \e^{y \over l}\sqrt{\left(\partial_v y\right)^2
+ {l^2 u_0^2 \over 4}\e^{2y \over l}}\ .
\ee
The Euler-Lagrange equation (\ref{W4}),
\bea
\label{W5}
0&=& {\e^{y \over l} \over 2}\sqrt{\left(\partial_v y\right)^2
+ {l^2 u_0^2 \over 4}\e^{2y \over l}} + {l^2 u_0^2 \e^{3y \over l}
\over 8\sqrt{\left(\partial_v y\right)^2
+ {l^2 u_0^2 \over 4}\e^{2y \over l}}} \nn
&& - \partial_v\left( {l\e^{y \over l}\partial_v y \over 2\sqrt{\left(\partial_v y\right)^2
+ {l^2 u_0^2 \over 4}\e^{2y \over l}}}\right)\ ,
\eea
shows that the following quantity is constant:
\be
\label{W6}
E={l^3 u_0^2 \e^{3y \over l} \over 8\sqrt{\left(\partial_v y\right)^2
+ {l^2 u_0^2 \over 4}\e^{2y \over l}}}\ .
\ee
If there is a turning point at $y=y_0$, where $\partial_v y=0$, then
\be
\label{W7}
\e^{2y_0 \over l}={4E \over l^2 u_0}\ .
\ee
Therefore, in order for  $y\geq y_0$ to be large,
${4E \over l^2 u_0}\gg 1$. Now $y$ is parameterized as
\be
\label{W8}
\e^{-{2y \over l}}= \e^{-{2y_0 \over l}}\sin \phi\ .
\ee
Then $\phi=0$ and $\phi=\pi$ correspond to the boundary $y\to \infty$, and
$\phi={\pi \over 2}$
to the turning point $y=y_0$. Combining (\ref{W6}), (\ref{W7}), and (\ref{W8}), one
finds
\be
\label{W9}
{dv \over d\phi}={1 \over u_0}\e^{y_0 \over l}\sqrt{\sin\phi}\ .
\ee
 From the metric (\ref{W3}), the distance $L_{q\bar q}$ between a quark and
an anti-quark
is given by
\be
\label{W10}
L_{q\bar q}=u_0\int dv = \e^{y_0 \over l}\int_0^\pi d\phi \sqrt{\sin\phi}
=\e^{y_0 \over l}{\sqrt{2}\Gamma\left({3 \over 4}\right)^2 \over \Gamma\left(
{3 \over 2}\right)}\ .
\ee
By using (\ref{W6}), (\ref{W7}), (\ref{W8}), and (\ref{W9}), the following
expression for the action (\ref{W4}) is obtained:
\be
\label{W11}
S_N={\Delta u l^2 \e^{y_0 \over l}\over 4}\int_0^{2\pi}d\phi \sin^{-{3 \over 2}}\phi\ .
\ee
The last expression, however, diverges at $\phi=0$ and $\phi=\pi$. The divergence
occurs because it contains the (infinite)  self-energy of the quark and
the anti-quark.
The self-energy can be evaluated by considering the configuration $0< y <
\infty$,  $x^i=0$, and $v$ is constant. Choosing $\sigma=y$ in the
Nambu-Goto action
(\ref{SFN20}), we obtain
\be
\label{W12}
S_{\rm self}={\Delta u l \over 2}\int_0^\infty \e^{y \over l}dy\ .
\ee
It is natural to divide the region of integration with respect to $y$ into
$0<y<y_0$ and $y_0\leq y < \infty$. In the region $y_0\leq y < \infty$, we parameterize
$y$ by (\ref{W8}). The action for the self-energy is
\be
\label{W12b}
S_{\rm self}=-{\Delta u l^2 \over 2}\left(1 - \e^{y_0 \over l}\right)
+ {\Delta u l^2 \over 8}\e^{y_0 \over l}\int_0^{\pi \over 2}d\phi
{\cos\phi \over \sin^{3 \over 2} \phi}  .
\ee
Then the action, subtracted by  the self-energy contribution, is given by
\bea
\label{W13}
S_N - 2 S_{\rm self} &=& {\Delta u l^2 \e^{y_0 \over l}\over 4}\left(4 +
{2\Gamma\left({3 \over 4}\right)^2 \over \Gamma\left(
{3 \over 2}\right)}\right) +  \Delta u l^2\left(1 - \e^{y_0 \over l}\right) \nn
&=& {\Delta u l^2 \over 2\sqrt{2} L} + \Delta u l^2 \ .
\eea
In the last expression, we used (\ref{W10}) and dependence on  $\e^{y_0
\over l}$ is absent.
The potential $V$ between the quark and anti-quark is then given by
\be
\label{W14}
V={l^2 \over 2\sqrt{2}L} + l^2 \ ,
\ee
and it does not depend on  the time coordinate $u$ or $u_0$.

Of course, the assumption that $u$ is almost constant
might not be a correct one  for a very small $u$.
However, it should  be valid for not too  small $u$.
This consideration shows that the Wilson loop (and possible confinement)
could be considered in time-dependent AdS/CFT.
The potential (\ref{W14}) is Coulomb-like potential except for
 the last term, which can
be re-absorbed into the self-energy of the quark and the anti-quark. The
Coulomb-like behavior in the potential  does not change from that
in the usual AdS/CFT case,  and  due to this term the quarks would not be
confined.  Thus,
the particle creation seems to affect the potential;
 it   seems to imply the same behaviour as that for usual AdS
 space.

\section{Discussion}

In summary, we discussed the properties of 5d cosmological AdS
spaces obtained from the   SAdS black hole cousins by  exchanging
the temporal  and spatial coordinates and employing an analytic
continuation when necessary.  There are similarities   between
such time-dependent spaces and black holes (cosmological horizon,
Hawking temperature).  An attempt to address the time-dependent
AdS/CFT correspondence, by accounting for the particle creation
effects and the presence of multiple vacua, is presented.  In
particular, we investigate in detail the role of surface
counter-terms in time-dependent AdS spaces. We evaluate quantities
which are identified with the surface energy-momentum tensor and
the (adiabatic) Wilson loop  on the gravity side (when AdS/CFT
correspondence is applicable). An explicit example of AdS
cosmological backgrounds which lead to the correct holographic
conformal anomaly
is presented. We  also addressed properties of a massive scalar
propagator  and found that the scaling dimensions of these fields
indicate an agreement with the dual field theory predictions.

The study of these specific  time-dependent backgrounds (as
solutions of string theory) indicates  that certain aspects of a
duality between such spaces and dual field theories in one
dimension less can be addressed. Of course,  it remains as  a
challenge to address an analog of such a duality within  the
realistic expanding Universe.

\section*{Acknowledgments}

The research  is supported in part by the Ministry of
Education,
Science, Sports and Culture of Japan under the grant n. 13135208
(S.N.),   RFBR grant 03-01-00105 (S.D.O.), LRSS grant 1252.2003.2
(S.D.0.), research funds from  ICREA (S.D.O.),  DOE grant
DOE-FG02-95ER40893 (M.C.), NATO linkage grant No. 97061
(M.C.), National
 Science Foundation Grant  No. INT02-03585 (M.C.) and the  Fay R. and
Eugene L.  Langberg Chair (M.C.).
 M.C. would like to
thank the  Institute for Advanced Study, Princeton,
 for hospitality and
support  during the course of this work.

\appendix

\section{Surface stress tensor and Wilson loop for T-dual backgrounds \label{s2}}

In this Appendix, we  investigate  properties of time-dependent
spaces obtained via T-duality   of the original AdS space-times.
 The obtained space-time is not,
however, asymptotically AdS. That is the reason why we relegated
this investigation to the Appendix. Nevertheless, as the starting
space was AdS,  there may be  remnants of the  holographic
correspondence for the T-dual space-time. We shall calculate
analogs of the surface stress tensor and Wilson loop and  argue
about a potential dual field theory interpretation.

When a $\sigma$ model in two dimensions has an Abelian or non-Abelian isometry,
one can introduce gauge fields by gauging the isometry and impose a constraint which
determines the gauge curvature to be  zero.
After imposing a gauge fixing condition and integrating out gauge fields, one obtains
a dual model.
In the dual model, there appears a shift of the dilaton fields when we use a
regularization which keeps the general covariance.

The non-Abelian $T$-duality \cite{quevedo} can be used to show
the equivalence of string models corresponding to
different topologies \cite{alvarez}.
Such a duality for the $SU(2)$ chiral model has been investigated in
\cite{zachos}. By using the $SU(2)$ $T$-duality, from the static 4d
space-time, a non-trivial space-time can be constructed like in \cite{SNcosm}.

The $T$-duality is a symmetry of the string theory. The low-energy
effective action (of the  Neveu-Schwarz Neveu-Schwarz sector) of
string theory  contains the dilaton field and the (rank two)
anti-symmetric tensor field. By the $T$-duality, not only the
metric but also the dilaton field and the anti-symmetric tensor
field are non-trivially modified.

Here we apply the $SU(2)$ $T$-duality to the Schwarzschild-AdS black hole (\ref{AdSS}).
One may write the metric $d\Omega_3^2$ in (\ref{AdSS}) by using the group element $g$
of $SU(2)$, which is topologically S$^3$.
\be \label{AFN1} d\Omega_3^2={1 \over 2}\tr g^\dagger d g
g^\dagger dg\ . \ee Here we parameterize $g$ by 3 parameters
$\phi_i$ ($i=1,2,3$): \be \label{coord} g=\e^{i\phi^i\sigma^i}\ .
\ee Here $\sigma^i$'s ($i=1,2,3$) are Pauli matrices. In general,
string theory can be formulated as the $\sigma$-model on the
2-dimensional Riemann surface. The target space of $\sigma$-model
corresponds to the real space-time, where string propagates. Then
by rewriting the $\sigma$-model in an equivalent way, one can
observe the $T$-duality of the string theory. Now  introducing the
2-dimensional $SU(2)$ gauge field as an auxiliary field, we
rewrite the $\sigma$-model.
By using $g$  (\ref{coord}), the Lagrangian density of the $\sigma$ model
corresponding to SAdS metric is given by
\be
\label{AFN1b}
{\cal L}_0 = {1 \over 2}\left\{ - f(r)\partial t \bar\partial t
+ {1 \over f(r)} \partial r \bar \partial r + {r^2 \over 2}
\tr g^\dagger \partial g g^\dagger \bar\partial g \right\}\ .
\ee
The above Lagrangian has $SO(4)\sim SU(2)_L \otimes SU(2)_R$ symmetry.
The two $SU(2)$ transformations are given by
\bea
&&g\rightarrow hg\ \ \ (SU(2)_L)\ ,\ \ \ \
g\rightarrow gh\ \ \ (SU(2)_R) \ .
\eea
($h$ is a group element of $SU(2)$).
In order to obtain a dual Lagrangian, we gauge
$SU(2)_L$ symmetry by introducing gauge fields $A$
and $\bar A$,
\bea
\label{gaugef}
&&A=A^iT^i\ ,\ \ \ \bar A=\bar A^iT^i \\
&&T^i={1 \over 2}\sigma^i\ ,
\eea
 replace
\be
\label{gauged}
\tr g^\dagger \partial g g^\dagger \bar\partial g \to
\tr g^\dagger \left(\partial + A\right) g g^\dagger
\left (\bar\partial + \bar A\right) g\ ,
\ee
and add a term which leads to vanishing of the gauge curvature
$F=[\partial + A, \bar\partial + \bar A]$,
\be
\label{constraint1}
\cL^{\rm constraint}={1 \over 2}\tr \xi F
\ee
Here $\xi$ is an element of $SU(2)$ algebra
\be
\label{constr}
\xi=\xi^i T^i\ .
\ee
If we first integrate $\xi$, we obtain a constraint that the gauge
curvature should vanish, which  implies that gauge fields are pure
gauge. Then one can choose the gauge condition that the gauge
fields vanish. The total Lagrangian including the terms
(\ref{constraint1}) and (\ref{constr}) is reduced to the original
Lagrangian (\ref{AFN1b}). On the other hand, the dual Lagrangian
can be obtained by integrating the gauge fields first.
In fact, when we integrate out the gauge fields $A$ and $\bar A$ first by
choosing the gauge condition
\be
\label{condition}
\phi^i=0\ ,
\ee
the dual Lagrangian is obtained as
\bea
\label{dual}
{\cal L}^{\rm dual} &=& {1 \over 2}\left\{ - f(r)\partial t \bar\partial t
+ {1 \over f(r)} \partial r \bar \partial r \right. \nn
&& \left. + {2 \over r^4 + 4 (\xi)^2}
\left[2r^2\delta_{ij}-4\epsilon_{ijk}\xi^k +{8 \over r^2}\xi^i\xi^j
\right]\partial\xi^i\bar\partial\xi^j\right\} \ .
\eea
Here $(\xi)^2=\xi^i\xi^i$.
Furthermore, by using the regularization which preserves
the general covariance, there is
a dilaton term in the dual theory \cite{quevedo},
\bea
\label{dilaton}
\cL^{{\rm dilaton}}&=&
-{1 \over 4\pi}R^{(2)}\Phi \nn
\Phi&=&\ln (r^4 + 4 (\xi)^2)\ .
\eea
Here $\Phi$ is the dilaton field.
As the dilaton field blows up when $r$ is large,  the space time
is asymptotically not AdS, but corresponds to the so-called
dilatonic vacuum .  In this region the
string theory is strongly coupled  and  the supergravity
description of the string theory  in this limit may be invalid. If
a holographic interpretation were still valid, the strongly
coupled limit  of the CFT should correspond to the classical limit
of the string theory, and  the supergravity description might
still be valid.  However, those are speculations and in the
following, the  calculation of  the trace anomaly from the gravity
side will be employed to test if this specific aspect of AdS/CFT
correspondence could be addressed.

Eq.(\ref{dual}) implies that the target space metric is given by
\be \label{dualmet} ds^2=-f(r) dt^2 + {1 \over f(r)}dr^2 + {2
\over r^4 + 4 (\xi)^2} \left[2r^2\delta_{ij} +{8 \over
r^2}\xi^i\xi^j \right]d\xi^i d\xi^j \ , \ee and the antisymmetric
tensor field also becomes non-trivial \be \label{AFN2} B_{\xi^i
\xi^j}= - {8 \epsilon_{ijk}\xi^k \over r^4 + 4 (\xi)^2}\ . \ee
Furthermore, the polar coordinates for $\xi^i$ may be introduced:
\be \label{AFN3} \left\{\begin{array}{rcl}
\xi^1 &=& \zeta \sin\theta \cos \varphi \\
\xi^2 &=& \zeta \sin\theta \sin \varphi \\
\xi^3 &=& \zeta \cos\theta \\
\end{array}\right. \ .
\ee
Then, dual metric is of the form:
\be
\label{AFN4}
ds^2=-f(r) dt^2 + {1 \over f(r)}dr^2 +
{4 \over r^2}d\zeta^2 + {4r^2 \zeta^2 \over r^4 + 4 \zeta^2}
\left(d\theta^2 + \sin^2 \theta d\varphi^2\right)\ .
\ee

We shall   now consider an  analytic continuation. By replacing
\be
\label{AFN5}
t\to i\tilde\phi \ ,\quad \varphi\to i\tilde t\ ,
\ee
 the following static metric is obtained
\be \label{AFN6} d s^2=- {4r^2 \zeta^2 \over r^4 + 4 \zeta^2}
\sin^2 \theta d\tilde t^2 + f(r) d\tilde\phi^2 + {1 \over
f(r)}dr^2 + {4 \over r^2}d\zeta^2 + {4r^2 \zeta^2 \over r^4 + 4
\zeta^2} d\theta^2\ . \ee On the other hand if we consider the
following analytic continuation: \be \label{AFN7} t\to i\tilde\phi
\ ,\quad \theta\to {\pi \over 2} + i\hat t\ , \ee one obtains the
cosmological (time-dependent) metric which is not asymptotically
AdS \be \label{AFN8} d s^2= -{4r^2 \zeta^2 \over r^4 + 4 \zeta^2}
d\hat t^2 + f(r) d\tilde\phi^2 + {1 \over f(r)}dr^2 + {4 \over
r^2}d\zeta^2+ {4r^2 \zeta^2 \over r^4 + 4 \zeta^2} \cosh^2 \hat t
d\varphi^2  \ . \ee
In order to avoid the orbifold singularity at
$r=r_H$, we need to impose the periodicity for the coordinate
$\tilde\phi$ in Eqs.(\ref{AFN5}) and (\ref{AFN7}): \be
\label{AFN9} \tilde\phi \sim \tilde\phi + {1 \over T_H}\ . \ee
Here $T_H$ is the Hawking temperature (\ref{hwk}). Then, naturally
$r$ may be restricted by (\ref{AFN10}). The singularity at $r=r_H$
does not appear in (\ref{AFN6}) or (\ref{AFN8}).

In the metric (\ref{dualmet}) when $r$ is large, we obtain
\be
\label{SFN1}
d s^2= {l^2 \over r^2}dr^2 -{r^2 \over l^2} dt^2 + {4 \over r^2}
\sum_{i=1}^3 \left(d\xi^i\right)^2 \ .
\ee
Then the surface with constant $r$ is flat but  the spatial
distance scales by the inverse power of $r$, which is different
from the case of usual (Schwarzschild)  Ads black hole. On the
other hand, in the large $r$ limit, the metrics (\ref{AFN6}) and
(\ref{AFN8}) behave as
\bea \label{SFN2} d s^2&=& {l^2 \over
r^2}dr^2 - {4\zeta^2 \over r^2} \sin^2 \theta d\tilde t^2 + {r^2
\over l^2} d\tilde\phi^2
+ {4 \over r^2}d\zeta^2 + {4 \zeta^2 \over r^2} d\theta^2\ ,\\
\label{SFN3} d s^2 &=& {l^2 \over r^2}dr^2 - {4 \zeta^2 \over r^2}
d\hat t^2 + {r^2 \over l^2} d\tilde\phi^2 + {4 \over r^2}d\zeta^2+
{4 \zeta^2 \over r^2} \cosh^2 \hat t d\varphi^2  \ . \eea
 The metrics of the surface with constant $r$ (\ref{SFN2}) and
(\ref{SFN3}) can be obtained from the metric corresponding to
(\ref{SFN1}) by the analytic continuations (\ref{AFN5}) and
(\ref{AFN7}), which do not include the radial coordinate $r$. It
is interesting that the surface metrics are also flat.

By putting a boundary with constant $r$, we may introduce the
surface terms: 
\bea 
\label{sf9} 
S_b&=&S_b^{(1)} + S_b^{(2)} \nn
S_b^{(1)} &=& {2 \over \kappa^2}\int d^4 x \sqrt{- g} \nabla_\mu
n^\mu  \nn S_b^{(2)} &=& - \int d^4 x \sqrt{- g_{(4)}}
\left(\eta_1 + \eta_2 R_{(4)}\right)\ . \eea Here $S_b^{(1)}$ is
the Gibbons-Hawking surface term and $n^\mu$ is the unit  vector,
(outward)  normal  to the boundary, which is now \be \label{SFN4}
n^r=\sqrt{f(r)}=\sqrt{{r^2 \over l^2} + 1 - {\mu \over r^2}}\
,\quad \left(\mbox{other components}\right)=0\ . \ee The constants
$\eta_1$ and $\eta_2$ in $S_b^{(2)}$ are determined so that the
total action becomes finite. Then the surface energy-momentum
tensor  can be derived by \cite{BK}
 \bea \label{sf13}
T^{mn}&=&T^{mn}_1 + T^{mn}_2\ ,\nn T^{mn}_1&=&{1 \over \kappa^2}
\left(2g_{(4)}^{mn} \nabla_\mu n^\mu - 2\nabla^m n^n\right)\ ,\nn
T^{mn}_2&=&-\eta_1 g_{(4)}^{mn} - 2\eta_2\left({1 \over 2}
g_{(4)}^{mn}R_{(4)} - R_{(4)}^{mn}\right)\ . 
\eea
(Of course, the interpretation of the above  quantity as an energy
momentum tensor is valid only  if AdS/CFT is  applicable.) Here
$g_{(4)}^{mn} $ is the metric induced on the boundary and the
curvatures $R_{(4)}$ and $R_{(4)}^{mn}$ are constructed with the
help of $g_{(4)}^{mn}$. The calculation of stress-energy  tensor 
goes in the standard way. When $r$ is large, $T^{tt}_1$ and $T^{tt}_2$ behave as
\bea
\label{SFN10}
T^{tt}_1&\sim& {6l \over \kappa^2 r^2}\left(1 - {l^2 \over 2r^2} + {1 \over r^4}
\left({3 \over 8}l^4 + {\mu l^2 \over 2} - {16 \over 3}\xi^2 \right)\right) \nn
T^{tt}_2 &\sim& {l^2\eta_1 \over r^2} + {\left( -\eta_1l^4 + 18 \eta_2 l^2
\right) \over r^4} + {\eta_1 \left(l^6 + \mu l^4\right) \over r^6} \ .
\eea
We choose $\eta_1$ and $\eta_2$ so that the leading and sub-leading terms in the
total $T^{tt}$ are canceled:
\be
\label{SFN11}
\eta_1=-{6 \over l\kappa^2}\ ,\quad \eta_2 = - {l \over 6\kappa^2}\ .
\ee
Then
\be
\label{SFN12}
T^{tt}\sim T^{tt}_1 + T^{tt}_2 = {l \over \kappa^2 r^6}\left( -3 \mu l^2 -32\xi^2
 -{15 \over 4}l^4\right)\ .
\ee
Introducing time-like unit vector $q_\mu$,
\be
\label{SFN13}
q_t=\sqrt{f(r)}\ ,\quad \mbox{other components}=0\ ,
\ee
the mass can be defined by
\be
\label{SFN14}
M = \lim_{r\to \infty}\int d^3\xi m(r)\ ,\quad m(r)\equiv \sqrt{g_{(4)}}
f^{1 \over 2}T^{\mu\nu}q_\mu q_\nu\ .
\ee
Since
\be
\label{SFN15}
g_{(4)}= {64 r^2 \over \left(r^4+4\xi^2\right)^2},
\ee
 for large $r$
\be
\label{SFN16}
\sqrt{g_{(4)}}f^{1 \over 2} \sim {\cal O}\left(r^{-2}\right)\ .
\ee
On the other hand, by using (\ref{SFN12}), one gets
\be
\label{SFN17}
T^{\mu\nu}q_\mu q_\nu \sim {\cal O}\left(r^{-4}\right)\ .
\ee
We find $\lim_{r\to \infty} m(r)\to 0$, which indicates that in
this case the mass  vanishes.
Originally the surface counter-term $S_b^{(2)}$ (\ref{sf9}) has been
introduced to
make the total action or mass finite. Eq.(\ref{SFN15}) seems to indicate
that the action might
be finite except for  the divergence coming from the infinite area of the
constant $r$
surface. In this case  the surface term  $S_b^{(2)}$ may  not be
necessary what is different from the case of usual AdS space. Namely, if
one takes $\eta_1=\eta_2=0$,
$T^{tt}=T^{tt}_1$ and $T^{\mu\nu}q_\mu q_\nu$ remain finite even in the
limit
$r\to \infty$:
\be
\label{SFN18}
T^{\mu\nu}q_\mu q_\nu \to {6 \over \kappa^2 l}\ .
\ee

 One may now
make an attempt to
compare   the above (analog of) surface stress tensor with the conformal
anomaly
on  the CFT side. When $r$ is large, the metric tensor and curvatures on
the
surface of constant $r$  are
\bea
\label{tt1}
&& ds_{(4)}^2 = \sum_{\mu,\nu=0}^3 g_{(4)\mu\nu}dx^\mu dx^\nu \sim
-{r^2 \over l^2}dt^2 + {4 \over r^2}\left(d\xi^i\right)^2\ ,\nn
&& R_{(4)\xi^i t \xi^j t}=0\ , \quad R_{(4)\xi^i \xi^j \xi^k \xi^l}
\sim {48 \over r^6}\left(\delta_{ik}\delta_{jl} - \delta_{il} \delta_{jk}\right)\ .
\eea
 From (\ref{t1}) it follows that the geometry of the surface is
$R^1 \times S^3$ and the radius of $S^3$ is ${\sqrt{3} \over r}$.
 Since the trace anomaly generally has the
following form:
\be
\label{OVII}
T^A=b\left(F_{(4)}+{2 \over 3} \Box R_{(4)}\right) + b' G_{(4)} + b''\Box R_{(4)}\ ,
\ee
we find $T^A = {\cal O}\left(r^{-6}\right)$. On the other hand, from
direct calculation it follows
\bea
\label{tt2}
{T_1}^m_{\ m} &=& {6 \over \kappa^2 l}\left\{ -2 - {2l^2 \over r^2} + {1 \over r^4}
\left( \mu l^2 + {3l^4 \over 4} + {16 l^4 \over 3}\right)
+ {\cal O}\left(r^{-6}\right)\right\} \nn
{T_2}^m_{\ m} &=& -4\eta_1 + {36\eta_2 \over r^2}
+ {\cal O}\left(r^{-6}\right) \ .
\eea
In this case the anomaly could not  be reproduced even if one chose
$\eta_1$ and $\eta_2$  appropriately.
This indicates that  additional  (higher derivatives) counter-terms are
necessary. It could be also that AdS/CFT correspondence does not occur.

The above discrepancy between (\ref{OVII}) and (\ref{tt2})  is due to the
fact that  the scaling for the spatial coordinates $\xi^i$ is different
from that
of the temporal coordinate $t$ for the dual model (\ref{dualmet}).
When the radial coordinate $r$ becomes large, the scale of the spatial
direction becomes small but the temporal one becomes large.
This discrepancy may be solved if we could consider the $T$-dual model in
the
$t$-direction,  as in Hull's type
IIB$^*$ model \cite{hull}. In this case the time component of the metric
changes from  $-f(r) dt^2$ in (\ref{dualmet}) to $- {dt^2 \over f(r)}$
and the metric takes the form:
\be
\label{dualmet2}
d s^2={1 \over f(r)}\left( - dt^2 + dr^2 \right) + {2 \over r^4 + 4 (\xi)^2}
\left[2r^2\delta_{ij} +{8 \over r^2}\xi^i\xi^j \right]d\xi^i d\xi^j \ ,
\ee
Then the scaling for the temporal coordinate coincides with that of
the spatial coordinates. With the   new  metric
 (\ref{dualmet2}) one can repeat the evaluation of analog of surface
stress-tensor.
When $r$ is large,  $T^{tt}_1$ and $T^{tt}_2$ behave as
\bea
\label{SFN10BB}
T^{tt}_1&\sim& {6r^2 \over \kappa^2 l^3}\left(1 + {3l^2 \over 2r^2} + {1 \over r^4}
\left({3 \over 8}l^4 - {3\mu l^2 \over 2} - {16 \over 3}\xi^2 \right)\right) \nn
T^{tt}_2 &\sim& {r^2\eta_1 \over l^2} +  \eta_1 + {18 \eta_2 \over l^2}
+ {- \mu \eta_1 + 18 \eta_2 \over r^2} \ .
\eea
We choose $\eta_1$ and $\eta_2$ so that the leading and sub-leading terms
in the
total $T^{tt}$ are canceled:
\be
\label{SFN11BB}
\eta_1=-{6 \over l\kappa^2}\ ,\quad \eta_2 = - {l \over 6\kappa^2}\ .
\ee
Then
\be
\label{SFN12BB}
T^{tt}\sim T^{tt}_1 + T^{tt}_2 = {l \over \kappa^2 l^3 r^2}\left( -6 \mu l^2 -32\xi^2
+ 6 l^4\right)\ .
\ee
On the other hand,  instead of (\ref{tt2})
\bea
\label{tt2BB}
{T_1}^m_{\ m} &=& {6 \over \kappa^2 l}\left\{ -4 - {l^2 \over r^2} + {16\xi^2  \over r^4}
+ {\cal O}\left(r^{-6}\right)\right\} \nn
{T_2}^m_{\ m} &=& -4\eta_1 - {36\eta_2 \over r^2}
+ {\cal O}\left(r^{-6}\right) \ .
\eea
Then with the same choice of $\eta_1$ and $\eta_2$ as in (\ref{SFN11BB}),
the leading and sub-leading terms in the total $T^m_{\ m}$ are canceled
\be
\label{DD1}
T^m_{\ m}={T_1}^m_{\ m} + {T_2}^m_{\ m}
= {96 \xi^2 \over \kappa^2 l r^4}\ .
\ee
The above expression  still depends on $\xi$, but if we set
 $\xi^2=0$, the result is consistent with (\ref{OVII}).
Physical meaning of the $\xi^i$-dependence in (\ref{DD1}) is not clear.
If Eq.(\ref{DD1}) is exact,  proposed holographic correspondence may hold
only on the
subspace,
which satisfies $\xi^2 =0$. Another possibility might be that our choice of
the boundary of the space-time may not be the correct one for the
investigation of  the  correspondence. Now the boundary has been
given by choosing $r$ to be
 constant, and is taken to infinity in the final
limit. In
this limit
 we find the geometry of the surface to be
$R^1 \times S^3$ and the radius of the $S^3$  to be ${\sqrt{3} \over r}$.
With this choice of the boundary,
the  boundary is independent of $\xi^i$.
Note that we can always choose the boundary so that its shape could be
different. Generally one may choose the  boundary  that  depends on the
coordinates
$\xi^i$. If the shape of the boundary is changed, the extrinsic curvature $2\nabla_m n_n$
and the intrinsic curvature $R_{(4)mn}$ of the boundary can be changed,
which may result in the change of the surface energy momentum tensor. If the difference
of the surface energy momentum tensor from the original parameterization of the boundary
could survive in the limit of $r\to\infty$, the expression of the holographic trace
anomaly should also change.  These observations  again indicate that there
are specific  modifications that are needed in the application of the
AdS/CFT correspondence to time-dependent backgrounds.

In the following we shall consider  the Wilson loop \cite{rey} as
determined on the gravitational side. Of course, this is a rather
formal calculation which may have Wilson line interpretation only
if the holographic correspondence is applicable. Nevertheless, we
use AdS/CFT terminology for simplicity.
 In the case  when the standard AdS/CFT correspondence is
 applicable, the Wilson line  describes the potential between a quark and
an
anti-quark.

We start with the Euclidean
signature space-time.
By  Wick-rotating ($t\to it$) the metric (\ref{AFN4}) one arrives at
\be
\label{SFN19}
d s_{\rm E}^2=g_{\mu\nu}dx^\mu dx^\nu = f(r) dt^2 + {1 \over f(r)}dr^2 +
{4 \over r^2}d\zeta^2 + {4r^2 \zeta^2 \over r^4 + 4 \zeta^2}
\left(d\theta^2 + \sin^2 \theta d\varphi^2\right)\ .
\ee
In order to consider the gravitational Wilson loop, we start with the
Nambu-Goto string action:
\be
\label{SFN20}
S_N={1 \over 2\pi}\int d\sigma d\tau \sqrt{\det\left(g_{\mu\nu}\left(x^\rho
\left(\sigma,\tau\right)\right)\partial_\alpha x^\mu \left(\sigma,\tau\right)
\partial_\beta x^\nu \left(\sigma,\tau\right)\right)}\ .
\ee
Here $\sigma$ and $\tau$ are the coordinates on the string world-sheet and
$\{\partial_\alpha, \ \partial_\beta\}$ correspond to the derivative with
respect to $\{\sigma,\  \tau\}$.

The starting configuration of the string world-sheet is
\be
\label{SFN21}
\tau = t\ ,\quad \sigma = \varphi\ ,\quad r=r(\sigma)\ ,
\quad \theta={\pi \over 2}\ ,\quad \zeta=\zeta_0\ (\mbox{constant})\ .
\ee
As it is a  static configuration, the obtained result remains the
same, including the case with the  T-dual metric
(\ref{dualmet2}).
The action (\ref{SFN20}) reduces to the form:
\be
\label{SFN22}
S_N={1 \over 2\pi}\int d\sigma d\tau \sqrt{ \left({dr \over d\sigma}\right)^2
+ V(r)}\ ,\quad
V(r)={4r^2 \zeta_0^2 f(r) \over r^4 + 4\zeta_0^2}\ .
\ee
Using the Euler-Lagrange equation obtained from the action (\ref{SFN20}),
 the following quantity is conserved on the string world-sheet:
\be
\label{SFN23}
E={V(r) \over \sqrt{ \left({dr \over d\sigma}\right)^2 + V(r)}}\ .
\ee
Let us assume that there is a turning point at $r=r_m$, where
 ${dr \over d\sigma}=0$,  and
\be
\label{SFN24}
E=\sqrt{V\left(r_m\right)}\ .
\ee
For the case that the boundary exists at $r=r_0$ and both of $r_0$ and $r_m$ are large,
we have
\be
\label{SFN25}
V(r)= {4\zeta_0^2 \over l^2}\left(1 + {l^2 \over r^2} + {\cal O}\left(r^{-4}\right)\right)\ .
\ee
Then Eq.(\ref{SFN23}) with (\ref{SFN24}) can be rewritten as
\be
\label{SFN26}
\left({dr \over d\sigma}\right)=4\zeta_0^2 \left({1 \over r^2} - {1 \over r_m^2}\right)\ .
\ee
When we consider the bulk space-time in the region where $r\leq r_0$, the
r.h.s. of (\ref{SFN26}) is negative although the quantity in the l.h.s. is
non-negative and thus in this regime Eq.(\ref{SFN26}) is inconsistent.
Thus, there is no non-trivial solution.
On the other hand  the region $r\geq r_0$ in the bulk space is not a usual one,
however it may correspond to  the $T$-dual theory, where the large scale  region
($r\gg 1$) is replaced  with the small scale region ($\sim \textstyle{1\over r}\ll 1$).
Then the  solution of (\ref{SFN26})
is:
\be
\label{SFN27}
r=r_m\sqrt{1 - {4\zeta_0^2 \over r_m^4}\sigma^2}\ .
\ee
Here the constant of integration is chosen so that $\sigma=0$ corresponds
to $r=r_m$. Then the string end-points, which  correspond to a quark and
an anti-quark,  exist at
\be
\label{SFN28}
\sigma=\sigma_\pm\equiv \pm {r_m^2 \over 2\zeta_0}\sqrt{1 - {r_0^2 \over r_m^2} }\ .
\ee
Using (\ref{SFN19}) and (\ref{SFN21}), the geodesic distance $L$ between
$\sigma_+$ and $\sigma_-$ is given by
\be
\label{SFN29}
L={2r_0 \zeta_0 \left(\sigma_+ - \sigma_-\right) \over \sqrt{r_0^4 + 4\zeta_0^2}}
= {2r_0 r_m^2 \sqrt{1 - {r_0^2 \over r_m^2}} \over \sqrt{r_0^4 + 4\zeta_0^2}}\ .
\ee
Since $r_m\sim r_0$ for large $r_0$, $V(r)$ in (\ref{SFN26}) is almost constant
on the string world-sheet:
\be
\label{SFN30}
V(r)\sim {4\zeta_0^2 \over l^2}\left(1 + {l^2 \over r_0^2}\right)\ .
\ee
As a result, using (\ref{SFN23}) the action (\ref{SFN22}) can be evaluated as
\be
\label{SFN31}
S_N={1 \over 2\pi}\int d\sigma d\tau {V \over E}
\sim {\zeta_0 \over \pi l}\left(1 + {l^2 \over 2r_0}\right){1 \over T_H} \left(
\sigma_+ - \sigma_-\right)\ .
\ee
Here $T_H$ is the Hawking temperature (\ref{hwk}), which comes from the
periodicity of the Euclidean time. By using (\ref{SFN29}), we may further
rewrite (\ref{SFN31}) as
\be
\label{SFN32}
S_N\sim {\zeta_0 L \over \pi l}\left(1 + {l^2 \over 2r_0}\right){1 \over T_H}
{\sqrt{r_0^4 + 4\zeta_0^2} \over 2r_0 \zeta_0 }\ .
\ee
The (would be gravitational side) potential between quark and anti-quark
is found by the analogy with usual AdS calculation as
\be
\label{SFN33}
E(L)={2\pi T_H}S_N={2L \zeta_0 \over l}\left(1 + {l^2 \over 2r_0}\right)
{\sqrt{r_0^4 + 4\zeta_0^2} \over 2r_0 \zeta_0 }\ ,
\ee
It is linear with respect to the geodesic distance. It indicates the
confinement of  quarks.

Instead of  configuration (\ref{SFN21}), we may consider another
configuration of subject to:
\be
\label{SFN34}
\tau = t\ ,\quad \sigma = \zeta\ ,\quad r=r(\sigma)\ ,
\quad \theta=\theta_0\ ,\quad \phi=\phi_0\ (\theta_0, \phi_0\ :\ \mbox{constant})\ .
\ee
Then by the similar calculations, we obtain the potential between the quark 
and the anti-quark, which is again linear with respect to the geodesic distance:
\be
\label{SFN37}
E(L)={r_0 \over l}\left(1 + {l^2 \over 2r_0}\right) L\ .
\ee
We have discussed an  analog of  Wilson loop on the supergravity side
that accounts for the particle creation effects in section \ref{s6}.

\end{document}